\documentclass[11pt]{article}

\usepackage{amsmath,amssymb,amsfonts}
\usepackage{graphicx}
\usepackage{bm}
\usepackage{hyperref}
\usepackage{cite}
\usepackage{geometry}
\usepackage{color}
\usepackage[authoryear,round]{natbib}
\usepackage{setspace}       					
\title{\textbf{Elastic waveform inversion for double-couple \\microseismic source estimation in \\vertically fractured transversely isotropic media}}

\author{
Ujjwal Shekhar$^{1,a}$\thanks{ujjwal.shekhar@uib.no}, 
Einar Iversen$^{1,a}$, 
Florin A. Radu$^{1,b}$, 
Inga Berre$^{1,b}$ and
Morten Jakobsen$^{1,a}$
}
\date{
$^{1}$Center for Modeling of Coupled Subsurface Dynamics, University of Bergen, Norway\\
$^{a}$Department of Earth Science, University of Bergen\\
$^{b}$Department of Mathematics, University of Bergen\\
\today
}
\begin{document}
\maketitle
\begin{abstract}
Accurate characterization of microseismic events during fluid injection in sedimentary formations is essential to mitigate environmental risks. The source mechanism for microseismic events related to a slip on a fault plane is given by a double-couple. Waveform inversion has emerged as a promising technique for estimating the moment tensor and the position vector of double-couple sources. In most applications of waveform inversion for the moment tensor of double-couple sources, the formation is typically assumed to be isotropic or, less frequently, transversely isotropic. Modification of the moment-tensor representation to account for anisotropy created by aligned vertical fractures in transversely isotropic formations has not been included while inverting microseismic waveform data. In this study on synthetic microseismic data, we present a waveform inversion algorithm that includes this modification, considering the formation in the focal region to be vertically fractured transversely isotropic (VFTI) and possessing orthorhombic symmetry. Since VFTI media lack rotational symmetry, no assumptions have been made about the orientation of the fault plane where the slip occurred. The moment tensor of double-couple sources is formulated in terms of the elastic parameters of the VFTI medium and geometrical parameters which are slip magnitude, slip angle, fault dip, and azimuth angle of the fault-normal. We assume that the subsurface velocity model has already been obtained by full waveform inversion of the controlled-source seismic data. Source inversion is treated as a local optimization problem, and we invert for the source location and the geometrical parameters. These geometrical parameters are more directly constrained by seismic data than the moment tensor components and offer geologically meaningful insights. This approach enhances microseismic monitoring in fractured formations and can be extended to more complex anisotropic media, such as monoclinic systems.

\end{abstract}

\textbf{Keywords:}
Double-couple, microseismicity, seismic anisotropy, waveform inversion.

\section{Introduction}

Sedimentary formations are targets for fluid injection and extraction activities, such as CO$_2$ sequestration and hydrocarbon production. Microseismic events may occur during these activities due to the reactivation of the critically stressed pre-existing faults. It is important to locate the microseismic event and estimate its moment tensor components to analyze the geomechanical environment in the subsurface. Depending on the focal mechanism for microseismic events, the moment tensor source can be categorized as a single-dipole, double-couple or explosion-type source \citep{sw}. Any general moment tensor can also be decomposed into an isotropic, a compensated linear vector dipole, and a double-couple part \citep{vav2005,vav2015}. The most common type of moment tensor is the double-couple source, which represents the force equivalent of shear faulting on a planar fault in isotropic media \citep{vav2015}. The double-couple mechanism for microseismic events has been analyzed in detail by Grechka (\citeyear{gre}). It is important to analyze the subsurface geology to understand where the double-couple mechanism can operate and where it cannot. For example, some clay, cementitious rocks, and salt bodies exhibit thixotropy. The deformation within such material is not elastic because the particles require a long time to reorganize \citep{mew}. The particles within it flow, and the material changes shape in response to the shear stress. Therefore, we should not expect any double-couple mechanism capable of triggering microseismicity. However, the deformation in a fractured sedimentary rock, such as shale, is close to (if not exactly) linearly elastic. In such formations, we can usually expect double-couple microseismic events.

In isotropic formations, it is straightforward to conclude that the source mechanism is that of a double-couple. This can be achieved, in theory, by analyzing the P-wave radiation pattern \citep{shea}, and in practice, by performing the linear waveform inversion of high-frequency seismograms \citep{sil}. There are many other techniques in earthquake seismology, as pointed out by Eaton (\citeyear{eat}) to characterize the source considering an isotropic earth model. However, in anisotropic formations, characterization of the double-couple source is a challenging task. It is known that in anisotropic formations, moment tensors of kinematically double-couple microseismic events exhibit non-double-couple focal mechanisms \citep{gre,ros,vav2005}; except for the case of strike-slip faulting. Boitz et al. (\citeyear{boi}) systematically investigated the effects of the Thomsen parameters \citep{thom} on seismic moments and their potency-tensor isotropic equivalent, and revealed that anisotropy can have a significant influence on the interpretation of the source mechanisms. Continuous fluid injection itself can be one of the possible causes of the azimuthal anisotropy of the treated formation \citep{greyas}. These arguments indicate that it is important to include anisotropic effects in the mathematical model of the double-couple source in elastic media.

Sedimentary rocks are transversely isotropic because they are formed as a result of the successive deposition of sediments in thin layers over a geological time period. Some sedimentary rocks are intrinsically transversely isotropic because of the aligned platy minerals that form these rocks. Another factor that contributes to seismic anisotropy, specifically azimuthal anisotropy in sedimentary rocks, is the existence of natural vertical fractures. The fractures are typically vertical due to the triaxial stress-state within the Earth's crust (Schoenberg and Helbig, \citeyear{sch}; also see figs. 1 and 2 of Shekhar, \citeyear{shek}). The combination of thin vertical fractures and fine horizontal layering gives rise to an effective medium known as the vertically fractured transversely isotropic (VFTI) medium \citep{sch}. Fractures affect the moment tensor radiation pattern \citep{gib}, and neglecting the fracture-induced seismic anisotropy in sedimentary rocks may lead to an incorrect estimation of the moment tensor components. Therefore, in this study, we formulate the expression for the moment tensor of the double-couple source in VFTI media. In the process of obtaining moment tensor elements, we introduce the parameters associated with the fault: slip magnitude, slip angle, fault dip, and azimuth of the fault-normal, which we refer to as the geometrical parameters. It is important to note that instead of orienting the normal to the fault plane to one of the anisotropic axes, a general fault-normal is considered. The fault plane on which the slip occurred does not necessarily coincide with any of the symmetry planes. This is important because VFTI media do not have rotational symmetry.

Having accurate velocity and density models is crucial for reliable estimation of the source location and focal mechanism in microseismicity. The full waveform inversion (FWI) is used extensively in the reconstruction of the velocity model \citep{mor,bro}. Time-lapse velocity changes in the subsurface velocity model can also be estimated by performing FWI of repeating seismic events \citep{kam}. An overview of the developments in FWI has been presented in Virieux and Operto (\citeyear{vo}), and an overview of issues encountered while performing FWI in anisotropic media is provided in Alkhalifah (\citeyear{alkh}). Using FWI of the controlled-source seismic data, one can obtain the subsurface VFTI model. The elastic parameters of the subsurface VFTI model can be computed from three fracture parameters known as fracture weaknesses and five independent elastic parameters of the background in which the fractures are embedded. The background, in which fractures are embedded, is transversely isotropic with a vertical axis of symmetry (VTI). Using the multi-media strategy \citep{oha}, it is possible to perform FWI in a multi-stage manner. This means that the elastic parameters of the VTI background can be obtained prior to performing FWI for fracture weaknesses. If a representative unfractured VTI background region exists, the five independent VTI parameters can also be reconstructed in that region using an efficient inverse scattering approach to perform FWI \citep{jako,xian}. Subsequently, spatially varying fracture weaknesses can be obtained in the region of interest using the cross-hole seismic FWI \citep{shekh}. Once the subsurface VFTI model has been reconstructed, one can proceed to estimate the parameters of the double-couple microseismic source.

Numerous techniques exist for double-couple source characterization and localization, but each of these techniques has their own advantages and disadvantages. Waveform inversion has, here too, emerged as one of the most attractive methods. In global seismology, waveform inversion has been frequently used to estimate the moment tensor of an earthquake source (see, for example, Tromp et al., \citeyear{trom}; Liu and Tromp, \citeyear{liu}; Kim et al., \citeyear{kim}). It has also been used to determine the source location and moment tensor of natural earthquakes in geothermal fields assuming a 1D velocity model \citep{mm}. The advantages of the waveform inversion over arrival-time picking \citep{gei}, wavefield extrapolation \citep{mcm} or back-propagating a seismogram \citep{gaj,art} and travel-time inversion \citep{tong} have been given in Huang et al. (\citeyear{hua}), Wang and Alkhalifah, (\citeyear{wan}), Wang et al., (\citeyear{wang}). Michel and Tsvankin (\citeyear{mic}) developed an elastic waveform inversion algorithm for anisotropic media to estimate the 2D velocity model along with the source parameters from microseismic data. To overcome the problem of non-linearity arising in FWI due to the unknown source location and source-time function, Wang and Alkhalifah (\citeyear{wan}) developed a source function-independent microseismic FWI. They designed an objective function by convolving reference traces with the observed and modelled traces to mitigate the effect of the source ignition time. Wang et al. (\citeyear{wang}) extended this work to the elastic medium. The gradient of the objective function in these works on microseismic imaging was computed with the adjoint-state method \citep{lio,ple}. The advancement in waveform-based source location methods can be found in the work of Li et al. (\citeyear{li}). Despite significant advancements in waveform-based microseismic inversion, existing methods often overlook the influence of fracture-induced anisotropy on the fault-plane stiffness and do not directly invert geometrical parameters in VFTI media.

This paper presents a waveform inversion algorithm that accounts for vertical fractures in the focal region and performs inversion of geometrical parameters in heterogeneous VFTI media. Additionally, we locate the double-couple microseismic source using elastic waveforms. The paper can be organized as follows. In Section 2, we present the expression for a general double-couple moment tensor in vertically fractured transversely isotropic (VFTI) media. In Section 3, we demonstrate how to formulate the forward and inverse problems. The Gauss–Newton method is applied to reduce the error between observed and modeled data. Instead of inverting for six moment-tensor components, we invert for four geometrical parameters. In Section 4, we provide a theoretical analysis on the suitable design of the acquisition set-up for the double-couple microseismic events. In Section 5, we implement the inversion algorithm. The developed numerical scheme is applied to a layered model. Finally, we test the sensitivities of different inverted parameters in case the actual model (layered model) is replaced by the FWI-generated model and the fracture-free model. We also examine the effects of insufficient receiver coverage and noise in the amplitude spectrum of the observed data on the accuracy of the inverted parameters.

In this work, we mainly use symbolic notation. In places where tensorial notation is used, the lower-case indices take values from 1 to 3, and the capital letter indices take values from 1 to 6, unless otherwise stated. Einstein summation convention applies to repeated indices.

\section{Double-couple moment tensor in vertically fractured transversely isotropic media}

In sedimentary formations, we often observe a set of thin aligned vertical fractures embedded in a background, which is transversely isotropic with a vertical axis of symmetry (VTI) due to the triaxial stress-state within Earth's crust. Vertical fractures combined with horizontal fine layering (VTI background) in the low-frequency limit result in an equivalent vertically fractured transversely isotropic (VFTI) medium (Fig. 1). 

\begin{figure}
\centering
    \includegraphics[width=4in]{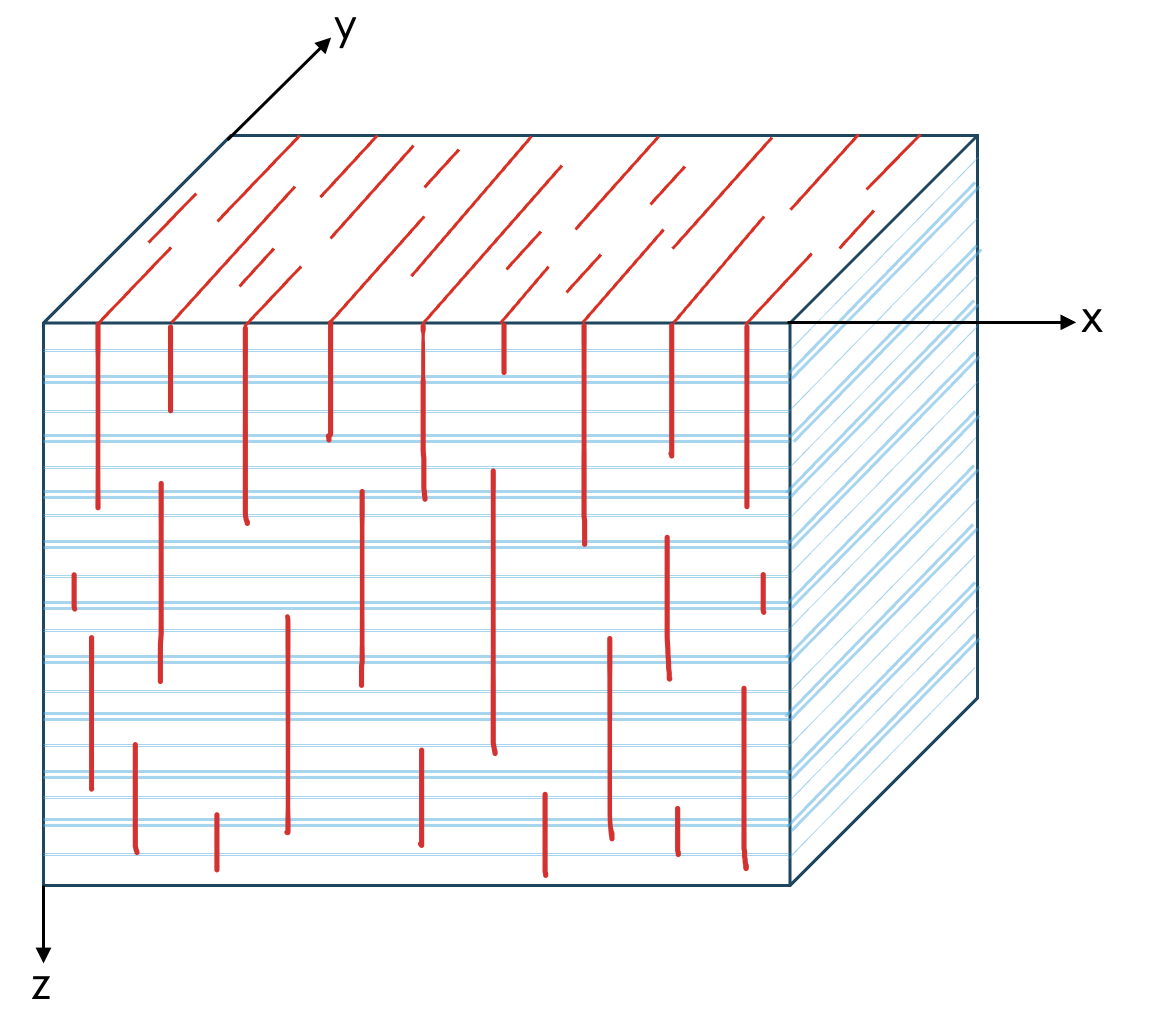}
    \centering
\caption{Vertically fractured transversely isotropic (VFTI) media with fracture planes parallel to yz-plane. Here, blue lines represent the horizontal layers and red lines represent the vertical fractures in the rock.}
\end{figure}

To obtain the moment tensor for a double-couple source, we need the stiffness matrix of the VFTI medium in the focal regime. The VFTI medium has orthorhombic symmetry. However, the total number of independent parameters in the stiffness matrix of VFTI media is eight instead of nine as in the case of general orthorhombic media. The stiffness tensor of the VFTI model is obtained following linear slip theory \citep{scho}. Let the stiffness tensor component (in Voigt notation) of a VTI background be $c^{(b)}_{IJ}({\bf x} )$. Using eqn. (10) of Schoenberg and Helbig (\citeyear{sch}), the nine orthorhombic-symmetry-type elastic parameters of a 6 x 6 symmetric stiffness matrix $c_{IJ}({\bf x} )$ of the VFTI medium can be expressed as 
\begin{equation}
\begin{aligned}
c_{11} &= c^{(b)}_{11}\left[1 - \delta_N\right], \qquad
c_{12} = c^{(b)}_{12}\left[1 - \delta_N\right], \qquad
c_{13} = c^{(b)}_{13}\left[1 - \delta_N\right],\\
c_{22} &= c^{(b)}_{11}\left[1 - \delta_N \frac{(c^{(b)}_{12})^2}{(c^{(b)}_{11})^2}\right], \qquad 
c_{23} = c^{(b)}_{13}\left[1 - \delta_N \frac{c^{(b)}_{12}}{c^{(b)}_{11}}\right], \quad\\
c_{33} &= c^{(b)}_{33}\left[1 - \delta_N \frac{(c^{(b)}_{13})^2}{(c^{(b)}_{11})(c^{(b)}_{33})}\right], \\
c_{44} &= c^{(b)}_{44}, \qquad 
c_{55} = c^{(b)}_{44}\left[1 - \delta_V\right], \qquad
c_{66} = c^{(b)}_{66}\left[1 - \delta_H\right] .
\end{aligned}
\end{equation}

In Eqn. (1), dimensionless quantities $\delta_N$, $\delta_V$, and $\delta_H$ are referred to as normal, vertical-tangential, and horizontal-tangential fracture weaknesses, respectively (Schoenberg and Helbig, 1997). Fracture weaknesses account for fracture density and filling material to describe the impact of fractures on seismic wave propagation. For more details on the fracture weaknesses and the theory of linear slip deformation used to obtain Eqn. (1), see Schoenberg and Helbig (1997) and appendix B of Shekhar et al. (\citeyear{shekh}). All quantities in Eqn. (1) may vary in space. For simplicity, the position vector {\bf x} associated with these quantities is omitted from the presentation.
 
 In a seismic rupture process involving a slip mechanism, the faulting is approximated as a double-couple of equivalent body forces \citep{sw}. Most of the time, complicated slip processes are approximated by a constant slip on a planar fault \citep{mad}. This approximation is needed to reduce the complications in computing the moment tensor elements. 
 
 Let the magnitude of the slip on the fault plane be $l$, the dip angle of the fault be $\alpha$, the azimuth angle of the fault-normal be $\theta$, and the angle between the slip vector and the dip direction on the fault plane be $\varphi$ (see Fig. 2). 
\begin{figure}
\centering
    \includegraphics[width=5in]{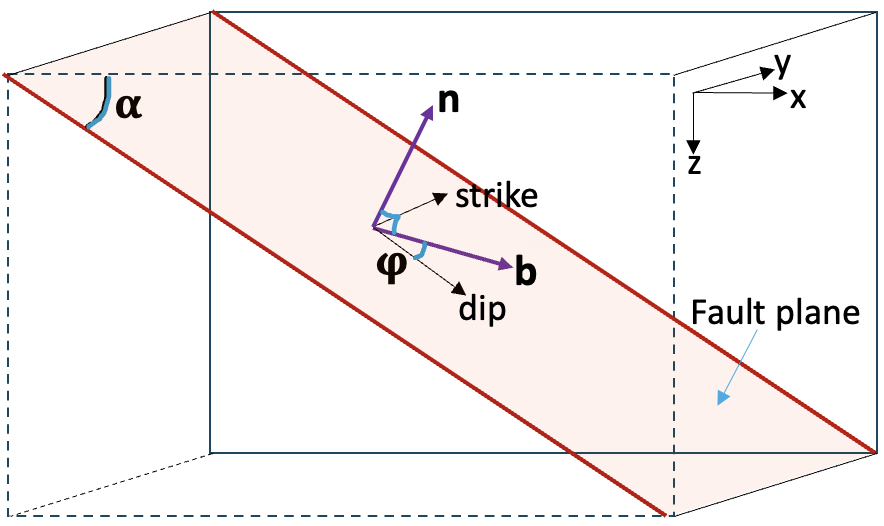}
    \centering
\caption{ Geometrical elements associated with the fault in a double-couple microseismic event. Here, ${\bf b}$ is the slip on the fault, ${\bf n}$ is the normal to the fault, $\alpha$ is the dip angle and $\varphi$ is the slip angle. Note that the strike of the fault plane can be in any direction and not necessarily parallel to y-axis.}
\end{figure}
Then the fault-normal ${\bf n}$ and the slip vector ${\bf l}$ are written as
 
 \begin{equation}
 {\bf n} = 
\left[
\begin{array}{c}
 \sin\alpha \cos\theta\\
 \sin\alpha \sin\theta \\
 \cos\alpha  
 \end{array}
\right],
 \end{equation} 
 and
  \begin{equation}
 {\bf l} = 
l\left[
\begin{array}{c}
\cos\alpha \cos\theta \cos\varphi- \sin\theta \sin\varphi \\
\cos\alpha \sin\theta \cos\varphi + \cos\theta \sin\varphi \\
 - \sin\alpha \cos\varphi 
 \end{array}
\right],
 \end{equation} 
 respectively. A second-rank potency tensor ${\bf D}$ that describes the kinematics of a seismic source (Boitz et al., 2018) is formulated as 
 \begin{equation}
 {\bf D} =  \frac{1}{2}\left[{\bf l} \otimes {\bf n} + {\bf n} \otimes {\bf l}\right],
 \end{equation} 
where $\otimes$ is the outer product of two vectors.

In general anisotropic media, the 3 x 3 moment tensor ${\bf M}$ representing the distribution of equivalent body force couples can be given as \citep{aki,gre}
\begin{equation}
 {\bf M} =  {\bf c}:{\bf D},
 \end{equation} 
 where {\bf c} is the fourth-rank stiffness tensor of the medium around the focal region and double-dot product ($:$) is the contraction of two tensors according to the last two indices of the first tensor and the first two indices of the second tensor \citep{auld}.
 The off-diagonal elements of the moment tensor ${\bf M}$ are symmetric.
 Using Eqns. (1-5) and eqn. (3) of Grechka (\citeyear{gre}), we obtain the moment tensor ${\bf M }$ of the double-couple source in VFTI media. It can be written as a function of the following parameters:
\begin{equation}
 {\bf M }= f\left(c^{(b)}_{IJ}, \delta_N, \delta_V, \delta_H, l, \alpha, \theta, \varphi \right). 
 \end{equation} 
The detailed expression for the elements of the moment tensor ${\bf M}$ is given as

 \begin{equation}
 \begin{aligned}
M_{11} &= l(1-\delta_N) \sin\alpha \biggl[ \biggl\{c^{(b)}_{11} + c^{(b)}_{12} - 2c^{(b)}_{13} + \left(c^{(b)}_{11} - c^{(b)}_{12}\right) \cos2\theta \biggr\} \frac{\cos \alpha \cos\varphi}{2} \\
& + \left(c^{(b)}_{12} - c^{(b)}_{11}\right) \sin\theta \cos\theta \sin\varphi\biggr] ,\\
M_{12} &= l(1-\delta_H) c^{(b)}_{66} \sin\alpha \biggl[ \cos\alpha \sin2\theta \cos\varphi + \cos2\theta \sin\varphi\biggr] ,\\
M_{13} &= l(1-\delta_V) c^{(b)}_{44}\biggl[ \cos2\alpha \cos\theta \cos\varphi + \cos\alpha \sin\theta \sin\varphi\biggr] ,\\
M_{22} &= l \sin\alpha \Biggl[ \Biggl\{c^{(b)}_{12}\left(1-\delta_N\right) + c^{(b)}_{11} \left(1- \delta_N\frac{c^{(b)2}_{12}}{c^{(b)2}_{11}}\right) - 2c^{(b)}_{13}\left(1- \delta_N\frac{c^{(b)}_{12}}{c^{(b)}_{11}}\right) \\
& + \biggl[c^{(b)}_{12}\left(1-\delta_N\right) - c^{(b)}_{11}\biggl(1- \delta_N\frac{c^{(b)2}_{12}}{c^{(b)2}_{11}}\biggr)\biggr] \cos2\theta \Biggr\} \frac{\cos\alpha \cos\varphi}{2} \\
& + \left\{c^{(b)}_{11} \biggl(1- \delta_N\frac{c^{(b)2}_{12}}{c^{(b)2}_{11}}\biggr) - c^{(b)}_{12}\left(1-\delta_N\right)\right\} \sin\theta \cos\theta \sin\varphi\Biggr] ,\\
M_{23} &= lc^{(b)}_{44}\biggl[ \cos2\alpha \sin\theta \cos\varphi + \cos\alpha \cos\theta \sin\varphi\biggr]  ,\\
M_{33} &= l \sin\alpha \Biggl[ \Biggl\{c^{(b)}_{13} \left(2-\delta_N - \delta_N\frac{c^{(b)}_{12}}{c^{(b)}_{11}}\right) - 2c^{(b)}_{33} \left(1- \delta_N\frac{c^{(b)2}_{13}}{c^{(b)}_{11}c^{(b)}_{33}}\right)  \\
& + \left(\frac{c^{(b)}_{12}}{c^{(b)}_{11}}-1\right)c^{(b)}_{13}\delta_N \cos2\theta \Biggr\}\frac{\cos\alpha \cos\varphi}{2} - \left(\frac{c^{(b)}_{12}}{c^{(b)}_{11}}-1\right)c^{(b)}_{13}\delta_N \sin\theta \cos\theta \sin\varphi\Biggr]. 
\end{aligned}
\end{equation}

\section{Modelling and inversion methods}

The double-couple microseismic source inversion in anisotropic elastic media is a highly non-linear problem because the location parameters and moment tensor components are coupled. In principle, we would like to obtain the source location, the moment tensor, the source signature, and the origin time from the waveform data. However, fully retrieving all these parameters simultaneously using an optimization method for this non-linear least squares problem is time-consuming and prone to the measurement error. The error can appear, for example, because of the uncertainty in the subsurface velocity model. Therefore, we should try to reduce the number of parameters that need to be determined. As mentioned by Huang et al. (\citeyear{hua}), the origin time directly causes a shift in arrival time for the observed seismogram, and can be combined with the source signature and be treated as a delayed source wavelet. Then, the source-function independent waveform inversion can be performed by using convolved wavefields (for more details, see Choi and Alkhalifah, \citeyear{choi}). Furthermore, Michel and Tsvankin (\citeyear{mich}) have shown that the wavelet can be estimated for a particular event from the elastic-waveform field data by stacking the traces along the event move-out. Thus, here we focus on determining the location and moment tensor components of the double-couple source. 

\subsection{Forward modelling}

The wave equation in anisotropic elastic media with the stiffness tensor ${\bf c}({\bf x})$ and the mass density $\rho ({\bf x})$ can be written in the velocity-stress formulation as
\begin{equation}
 \rho({\bf x})\frac{\partial {\bf v}({\bf x},t)}{\partial t } = \nabla\cdot \boldsymbol\tau ({\bf x},t) + {\bf f }({\bf x},t), 
\end{equation}  
\begin{equation}
 \frac{\partial \boldsymbol\tau ({\bf x},t)}{\partial t} = {\bf c}({\bf x}):\nabla_s {\bf v}({\bf x},t), 
\end{equation}  
where ${\bf x} \in \mathcal{R}^3$ is the space coordinate location, ${\bf v}$ is the particle velocity at time $t$, $\boldsymbol\tau$ is the stress tensor and ${\bf f }$ is the external force source density. Here, $\nabla_s$ represents the symmetric gradient of any physical quantity.

The equivalent body force density for the moment tensor source \citep{aki} is
\begin{equation}
 {\bf f}({\bf x}, t) = -{\bf M}\cdot\nabla \delta ({\bf x} - {\bf s})S(t),   
\end{equation}
where ${\bf s}$ represents the positioning of the double-couple microseismic event, $S(t)$ is the source-time function and $\delta({\bf x} - {\bf s})$ is the 3D Dirac-delta function. The equivalent body force density is implemented by evenly distributing the stress increments around the source location \citep{shi}. It should be noted that we multiply the body force by the volume of the grid block while discretizing. The fourth-order finite difference scheme over a staggered grid \citep{vir} is used to solve the elastodynamic wave equation for the particle velocity components. This forward solver with the perfectly matched layers (PML) boundary condition is implemented in MATLAB using the adaptation of codes from Lei et al. (\citeyear{lei}).

The particle displacement vector ${\bf u}$ can be computed from the particle velocity as
 \begin{equation}
 {\bf u} ({\bf x},\mathcal{T}) = \int_0^\mathcal{T} {\bf v} ({\bf x},t)dt,
\end{equation} 
where $\mathcal{T}$ is the time instant at which we obtain the particle displacement.

\subsection{Inversion}

The source parameters ${\bf m}$ in block matrix form can be given as a combination of the source location parameters ${\bf s}$ and the moment tensor parameters ${\bf A}$ as
\begin{equation}
 {\bf m}  \equiv \left[
\begin{array}{c}
 {\bf s} \\
  {\bf A} \\
 \end{array}
\right] ; \quad  {\bf A}  = \left[
\begin{array}{c}
 l \quad \alpha \quad \varphi \quad \theta
 \end{array}
\right]^{\mbox{\scriptsize T}}.
\end{equation}

To invert ${\bf m}$ from the elastic waveform data, the objective function (or misfit function) should be minimized. The objective function $\mathcal{E}$ is given by the L2-norm of the error between the simulated particle displacements ${\bf u}$ and the observed/recorded particle displacements ${\bf d}$ normalized with respect to ${\bf d}$. We write
\begin{equation}
\mathcal{E}\left[{\bf m}\right] = \frac{1}{2} \frac{\Vert {\bf r} \Vert_2}{\Vert {\bf d} \Vert_2},
\end{equation}
where
\begin{equation}
{\bf r} = \biggl({\bf u}\left[{\bf m}\right] - {\bf d}\biggr)
\end{equation}
is referred to as the data residual. Therefore, the objective function can be interpreted as the relative data error.

\subsubsection{Gradient and Hessian computation}

To compute the gradient and Hessian at various iterations of the inversion algorithm, we shall first formulate the Jacobian. The entries of the Jacobian ${\bf J}$ are the partial derivatives of the data residual ${\bf r}$ with respect to source parameters ${\bf m}$, and are given as 
\begin{equation}
 {\bf J}  \equiv
 \frac{\partial {\bf r}}{\partial {\bf m}} .
\end{equation}

The Jacobian related to the x, y and z coordinates of the source position can be represented by ${\bf J}^{s_1}$, ${\bf J}^{s_2}$ and ${\bf J}^{s_3}$ respectively; and the Jacobian related to the moment tensor parameters, i.e., slip $l$, dip angle $\alpha$, slip angle $\varphi$ and the fault-normal's azimuth angle $\theta$ can be represented by ${\bf J}^{l}$, ${\bf J}^{\alpha}$, ${\bf J}^{\varphi}$ and ${\bf J}^{\theta}$ respectively. Then,
\begin{equation}
 {\bf J}  \equiv \left[
\begin{array}{c}
 {\bf J}^{s_1} \quad {\bf J}^{s_2} \quad {\bf J}^{s_3} \quad {\bf J}^{l} \quad {\bf J}^{\alpha} \quad {\bf J}^{\varphi} \quad {\bf J}^{\theta}
 \end{array}
\right]^{\mbox{\scriptsize T}}.
\end{equation}

We find the Jacobian by perturbing each source parameter and computing the particle displacement at those perturbed values, followed by a finite-difference approximation of the partial derivatives. This means we have to perform two microseismic wavefield modelling simulations to obtain the Jacobian corresponding to one parameter. The change in the data residual is solely governed by the change in the simulated data ${\bf u}$, and therefore the Jacobian depends only on ${\bf u}$.
If there are $Nr$ receivers, $Np$ parameters of the double-couple source, and $Nt$ time steps at which particle displacements are recorded, then the size of the Jacobian ${\bf J}$ becomes $3Nr Nt$ by $Np$. It can be rearranged as a long vector of size $3Nr Nt Np$ by $1$. It should be noted that a factor of $3$ appears in multiplication because we are using all 3-components of the particle displacements at receivers.

Using the Jacobian, we compute the gradient ${\bf g}$ and the approximate Hessian ${\bf H}$, respectively as
\begin{equation}
 {\bf g}  = {\bf J}^{\mbox{\scriptsize T}} {\bf r},
\end{equation}
and
\begin{equation}
 {\bf H}  = {\bf J}^{\mbox{\scriptsize T}} {\bf J}.
\end{equation}

\subsubsection{Local optimization}

Let the parameters of the double-couple source in the $k$-th iteration of the waveform inversion routine be ${\bf m}^{(k)}$. Then the Gauss-Newton optimization algorithm to minimize the objective function $\mathcal{E}$ in Eqn. (13) is given as \citep{nw} 
\begin{equation}
{\bf m}^{(k+1)} = {\bf m}^{(k)} - {\bf H}^{-1}{\bf g}.
 \end{equation} 
The Gauss-Newton method provides rapid local convergence and has a number of advantages over other gradient-based methods (see Nocedal and Wright, 2000). We can rewrite Eqn. (19) as
\begin{equation}
{\bf H}\Delta {\bf m}^{(k)} = -{\bf g},
 \end{equation} 
where 
\begin{equation}
\Delta {\bf m}^{(k)} = {\bf m}^{(k+1)} - {\bf m}^{(k)}.
 \end{equation} 
The approximate Hessian takes into account the interaction between different parameters during the optimization. 

Double-couple microseismic source parameters' estimation can be challenging because of the inherent non-uniqueness and ill-posed nature of the inverse problem. Therefore, we add a regularization term \citep{men} and Eqn. (20) is formulated as
\begin{equation}
({\bf H} + \lambda^2 {\bf I})\Delta {\bf m}^{(k)} = -{\bf g},
\end{equation}
where ${\bf I}$ is the identity matrix, and $\lambda$ is the regularization term given by the trace of the Hessian matrix (Menke, 2012). The regularization term $\lambda$ gradually decreases with each inversion iteration according to the Levenberg–Marquard technique \citep{ast,jako}. 

To monitor the progression of the inversion algorithm with each iteration of numerical tests, the relative position error $\epsilon_{\bf s}$ is defined as
\begin{equation}
  \epsilon_{\bf s} = \frac{\lVert {\bf s} - {\bf st}\rVert_2}{\lVert {\bf st}\rVert_2}.
\end{equation}
Here, ${\bf s}$ is the updated location of the source at each inversion iteration, and ${\bf st}$ is the actual location of the source. The optimization algorithm is terminated when the objective function $\mathcal{E}$ reaches a prescribed minimum.

In case we invert for the moment tensor components directly, it is necessary to introduce the scaling coefficients corresponding to different parameters (see, for example, Michel and Tsvankin, \citeyear{tsv}). However, inverting for the fault and slip parameters (geometrical parameters) does not require such scaling, as the orders of their magnitudes (in their respective SI units) are not very different.

\section{Design of the acquisition geometry for source inversion}

The radiation pattern of the double-couple moment tensor in fractured media is considerably different from the patterns of isotropic and single-dipole seismic sources and other sources that can be represented by a vectorial point source (see Gibson and Ben-Menahem, \citeyear{gib}). For the double-couple, the perturbations in different geometrical parameters affect its radiation patterns. The direction of maximum energy of waves can change significantly with the variation in the dip of the fault and the direction of slip on the fault plane (see Eisner and Stanek, \citeyear{es}, for a detailed description of the directivity of microseismic events). The elastic energy distribution differs greatly in different directions. The borehole array may fall within the zone where the sensor receives weak amplitude or even no signals. The P and S-wave signals in surface microseismic data are usually contaminated with strong background noise \citep{cham,dun}, which means data recorded in the surface sensor need a lot of processing before being used for the inversion. When event azimuths cannot be derived from the P-wave polarization vectors, observations from two or more wells are required to locate the events and build azimuthally anisotropic velocity models \citep{yas}. This is especially necessary if the media possess lower-symmetry anisotropies such as monoclinic and triclinic. Vavryčuk (\citeyear{vav2007})) did an in-depth study on the retrieval of moment tensors from borehole data. He concluded that the complete moment tensor can be retrieved from amplitudes of P-waves provided that receivers are deployed in three parallel boreholes or from amplitudes of both P- and S-waves by deploying receivers in two non-parallel boreholes. Taking into account all these factors, it can be deduced that having multiple borehole arrays is important for the accurate determination of the source parameters. The boreholes must be at different azimuths from the source \citep{vav2007}. In general, to invert for the moment tensor components using the microseismic data, a necessary condition is that the receiver geometry samples a sufficiently large solid angle for the inverse problem to be well-posed \citep{eatfor}. In heterogeneous media, the strain field in heterogeneity creates a contrast source of the moment-tensor type. Depending on the strength of this virtual source, the inversion scheme may fail to locate the true source. It will also fail to determine the accurate geometrical parameters of the source. Therefore, we want to have more direct waves in our recording, which is mostly possible by using borehole receivers.

\section{Numerical results and discussion}

We estimate the position vector and geometrical parameters of the double-couple microseismic source using elastic waveforms recorded in vertically fractured transversely isotropic (VFTI) media. Throughout this section, whenever we mention terms ``source" and ``event", we mean ``double-couple microseismic source" and ``double-couple microseismic event", respectively. We perform all the numerical simulations for a 3D Cartesian mesh. The grid is uniformly spaced in each direction. The volume of each cell is the product of the grid spacing in the x, y and z directions. Each node characterizes a unique stiffness matrix and a density value. For simplicity, a single source is considered. However, generalization to multiple sources or a cluster of sources along a fault plane is straightforward and can be achieved by adding incremental stress contributions from each of the sources. The stiffness matrix ${\bf c}$ (in GPa) at the source position (derived following Schoenberg and Helbig, 1997) used to obtain the radiation patterns in VFTI medium is 
\begin{equation}
{\bf c} = 
\begin{bmatrix}
21.62 & 8.65  & 5.40 & 0 & 0 & 0\\
  & 22.78 & 5.61 & 0 & 0 & 0 \\
 &  & 13.71 & 0 & 0 & 0 \\
 &  & &  4.60 & 0 & 0\\
  &  & &  & 4.14 & 0 \\
  &  &  & &  & 5.86
\end{bmatrix}.
\end{equation}
In the true source model, the dip angle $\alpha$ is $\pi/4$ rad, the slip angle $\varphi$ is $\pi/3$ rad, and the magnitude of the slip is 1 m. The azimuth angle $\theta$ of the fault-normal is $\pi/4$ rad. A high-frequency Brune wavelet is used to represent the source-time function (Fig. 3) for the double-couple source because it demonstrates the sudden jump in the energy released in double-couple events and a typical spectral decay at high frequencies \citep{bru}. It also exhibits a skewed amplitude distribution, which is expected for mechanical deformation during microseismic activities.
\begin{figure}
\centering
    \includegraphics[width=3in]{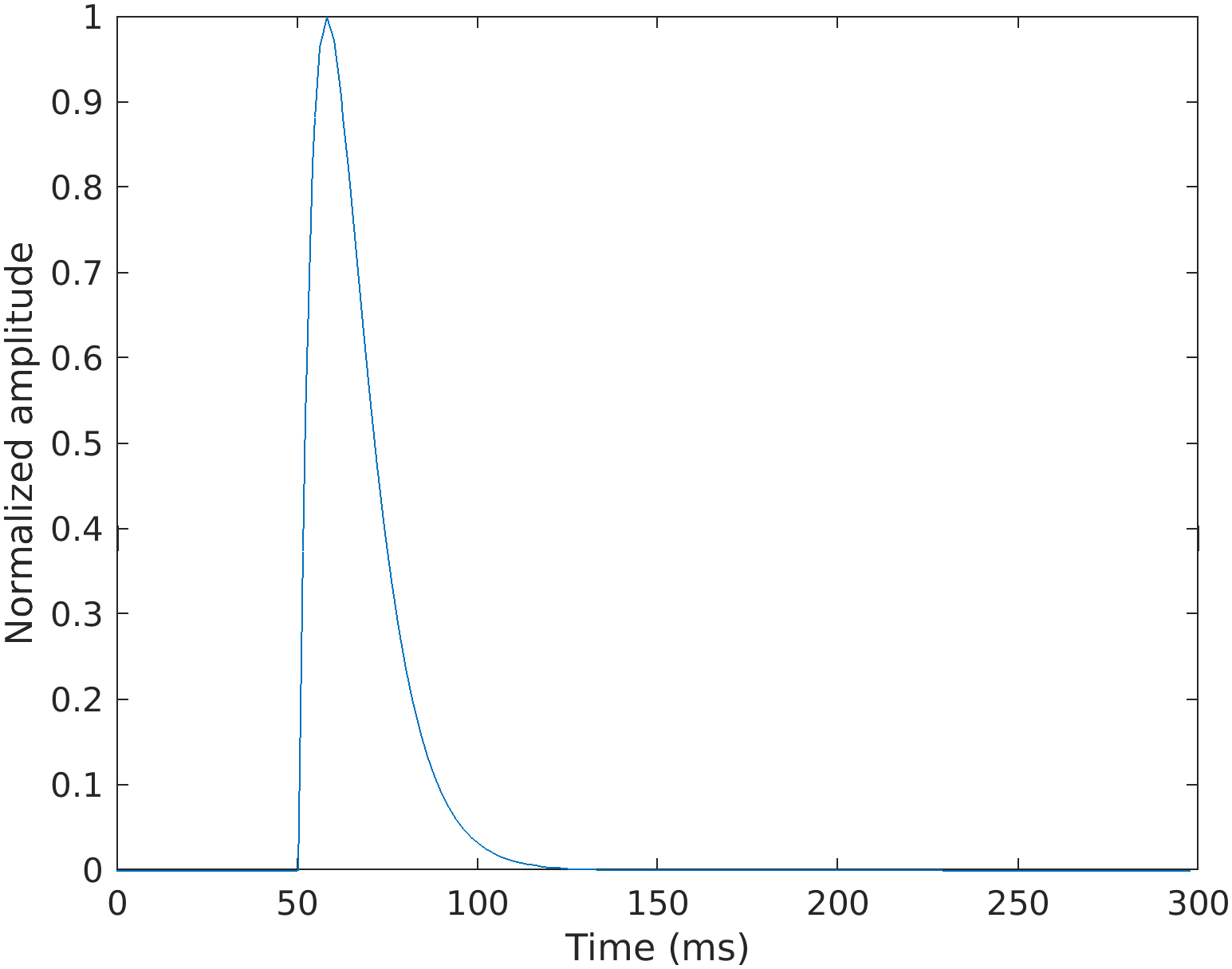}
\caption{The source-time function used to generate the synthetic particle displacement data.}
\end{figure}

Data from seismic surveys conducted in the region monitored for microseismicity should be utilized to build the anisotropic velocity model of the subsurface. We assume here that the velocity model is known or can be constructed separately by performing elastic FWI on controlled-source seismic data. The subsurface velocity model is a layered model that represents heterogeneous VFTI media.

\textbf{Layered model (the actual model)}

\begin{figure}
\centering
    \includegraphics[width=3.5in]{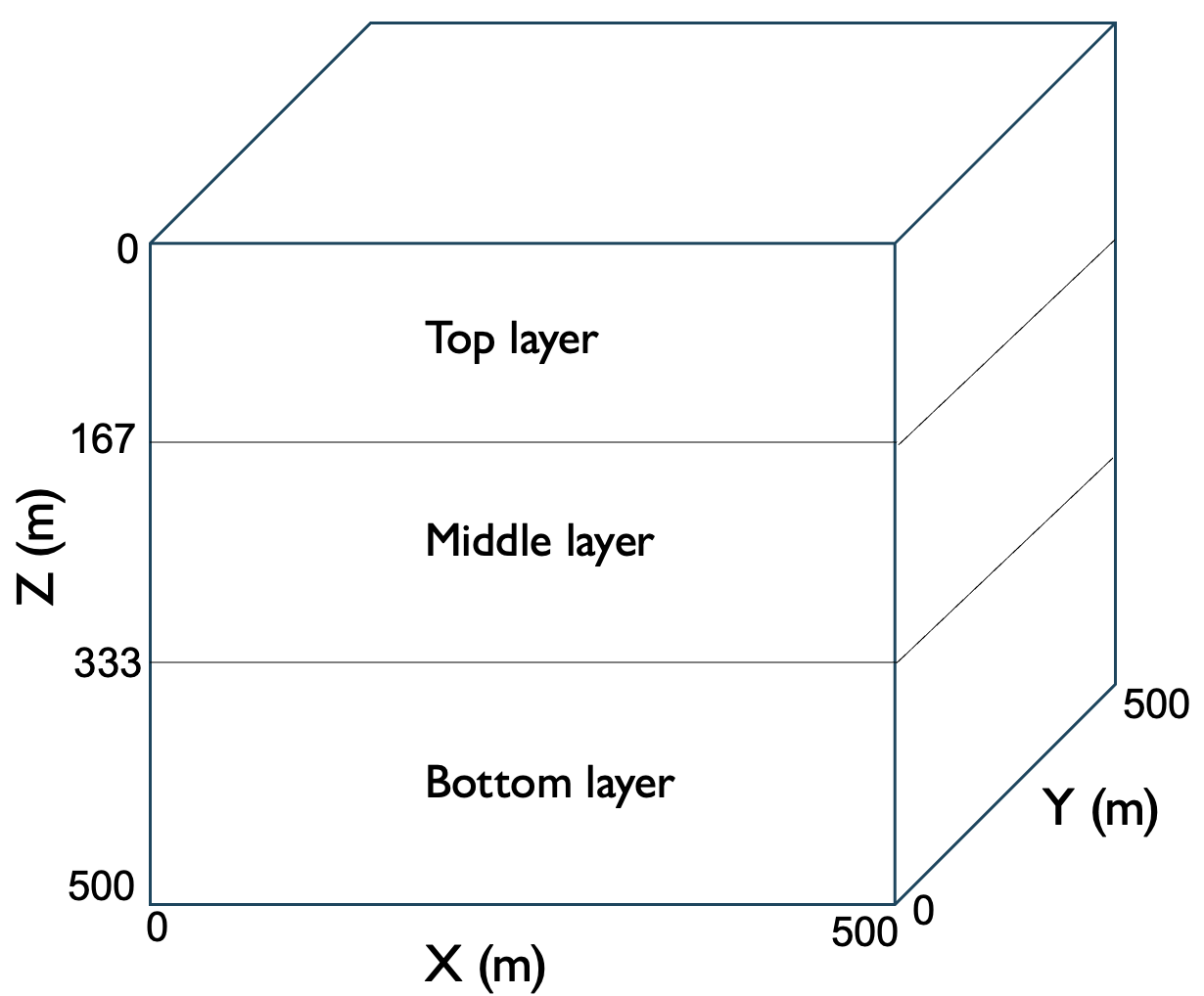}
    \centering
\caption{ The layered model is a typical representation of the overburden-reservoir-underburden system, and is used to test the source inversion scheme. The elastic parameters, fracture weaknesses and densities values in different layers are given in Table 1.}
\end{figure}
The layered model has three layers (Fig. 4). The dimensions are 500 m in each direction, and the grid size is 4 m. The grid size should be one-fourth (or less) of the wavelength of the slowest S wave in the model. The time step is chosen to satisfy the Courant–Friedrichs–Lewy condition. The stiffness coefficients of the VFTI medium (see Eqn. 1) consist of the stiffness coefficients $c^{(b)}_{IJ}$ of the unfractured VTI background (in which fractures are embedded) and a set of fracture weaknesses (Schoenberg and Helbig, 1997). The values of the elastic parameters $c^{(b)}_{IJ}$, the fracture weaknesses, namely the normal ($\delta_N$), the vertical-tangential ($\delta_V$) and the horizontal-tangential ($\delta_H$) fracture weaknesses, and the density $\rho$ in different layers are given in Table 1.

\begin{table}[h!]
\centering
\begin{minipage}{175mm}
\caption{The elastic parameters $c_{IJ}^{(b)}$ (in units of GPa), the fracture weaknesses $\delta_N$, $\delta_V$ and $\delta_H$, and the density $\rho$ (in units of kg/m$^3$) in the layered model}
\label{anymode}
\begin{center}
\begin{tabular}{@{}llllll}
\hline
\hline
Stiffness coefficients & Top layer & Middle layer & Bottom layer \\
\hline
$c_{11}^{(b)}$ & 22.50 & 23.00 & 23.50  \\[2pt] 
$c_{33}^{(b)}$ & 13.50 & 13.80 & 14.10  \\[2pt]
$c_{44}^{(b)}$ & 4.50 & 4.60 & 4.70  \\[2pt]
$c_{66}^{(b)}$ & 6.75 & 6.90  & 7.05 \\[2pt]
$c_{13}^{(b)}$ & 5.62 & 5.75 & 5.88  \\[2pt]
$\delta_N$ & 0.10 & 0.06 & 0.00  \\[2pt]
$\delta_V$ & 0.15 & 0.10 & 0.00  \\[2pt]
$\delta_H$ & 0.20 & 0.15 & 0.00  \\[2pt]
$\rho$ & 2250 & 2300 & 2350  \\[2pt]
\hline
\hline
\end{tabular}
\end{center}
\end{minipage}
\end{table}

\subsection{Wavefield modelling}

Since the peak frequencies of direct waves for microseismic events are high, fine sampling in space and time is required to calculate wavefields within the model. The magnitude of x-, y- and z- components of the elastic displacement field $u_i^{\mbox{norm}}$, at time $t$, at any position ${\bf x}$ within the model are given by the respective particle displacement components $u_i$ at ${\bf x}$ normalized with respect to the maximum amplitude of the corresponding particle displacement component in the whole model. Let the maximum amplitude $U_i$ at time $t$ be observed at ${\bf x}'$. Then
\begin{equation}
  u_i^{\mbox{norm}}({\bf x},t) = \frac{ u_i({\bf x},t) }{\vert U_i({\bf x}',t) \vert}; \quad  t > t_0,
\end{equation}
where $t_0$ is the timing of initiation of the event.
\begin{figure}
 \hspace*{-0.5in}
{(a)}
\begin{minipage}{0.30\textwidth}
    \includegraphics[width=2.7in]{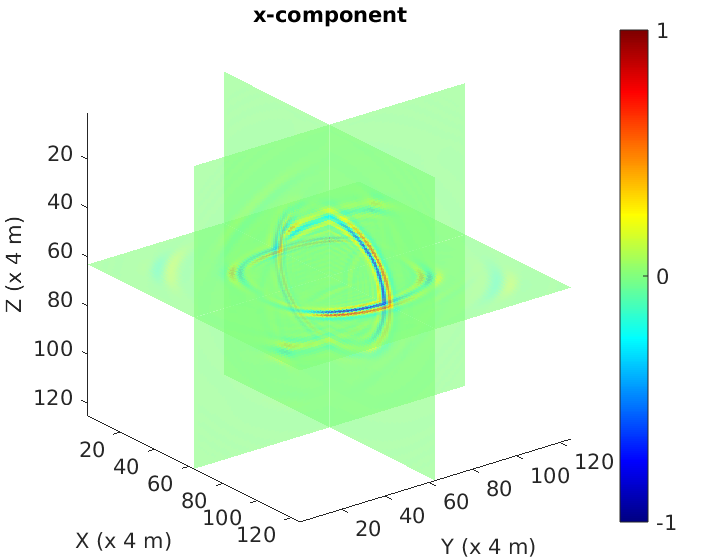}
    \centering
\end{minipage}
\hspace*{0.37in}
\begin{minipage}{0.30\textwidth}
    \includegraphics[width=2.7in]{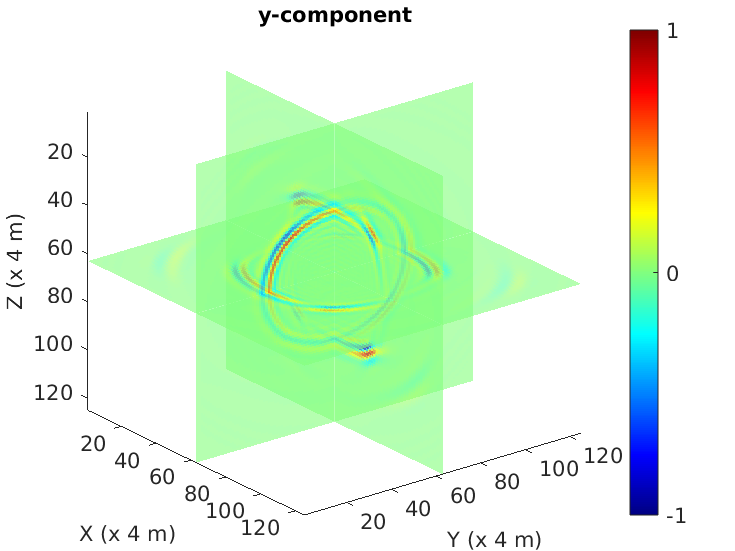}
    \centering
 \end{minipage}
\hspace*{0.37in}
 \begin{minipage}{0.30\textwidth}
    \includegraphics[width=2.7in]{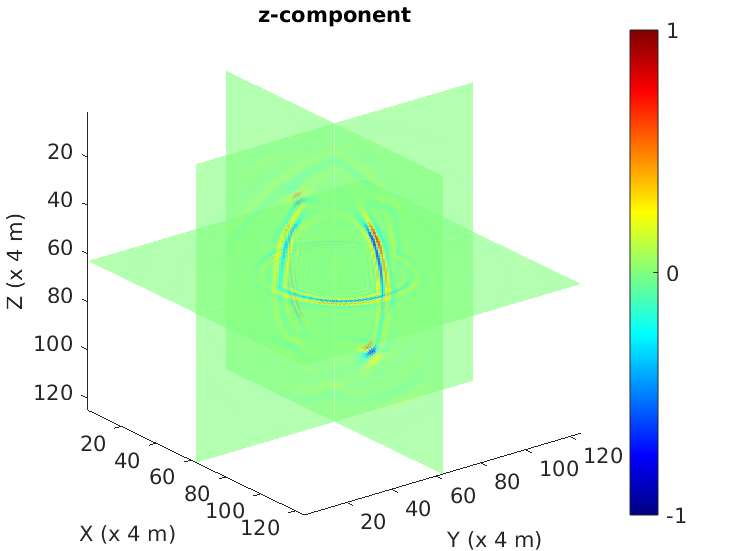}
    \centering
\end{minipage}\\[1cm]
 \hspace*{-0.5in}
{(b)}
\begin{minipage}{0.30\textwidth}
    \includegraphics[width=2.7in]{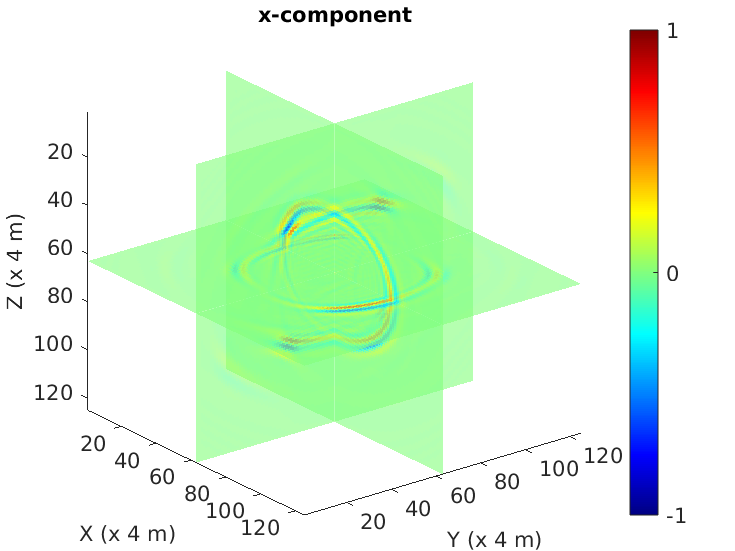}
    \centering
 \end{minipage}
 \hspace*{0.37in}
 \begin{minipage}{0.30\textwidth}
    \includegraphics[width=2.7in]{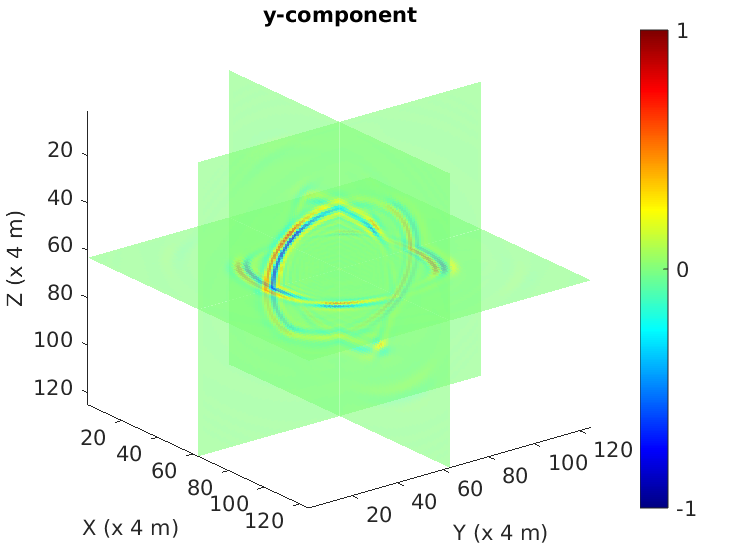}
    \centering
\end{minipage}
\hspace*{0.37in}
\begin{minipage}{0.30\textwidth}
    \includegraphics[width=2.7in]{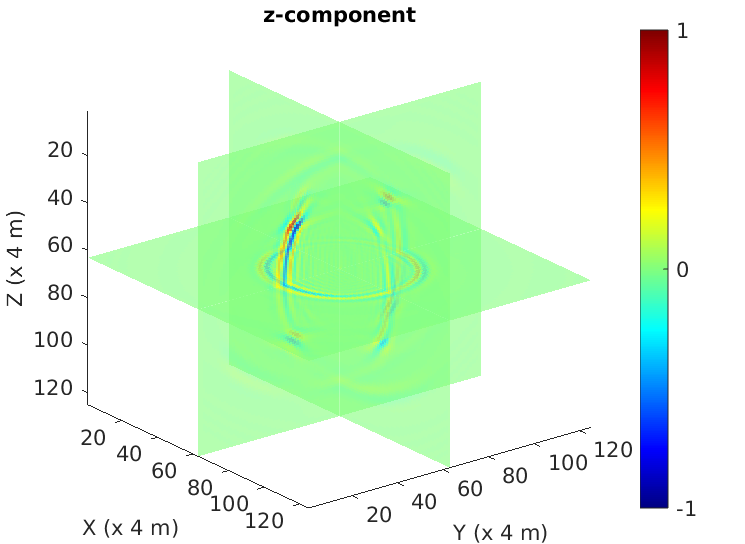}
    \centering
 \end{minipage}\\[1cm]
\caption{Snapshot of the elastic displacement field due to the double-couple source at the center of the layered model when the azimuth angle $\theta $ of the fault-normal is (a) $\pi/6$ rad, and (b) $2\pi/3$ rad.}
\end{figure}

In Fig. 5, we show the snapshot of the particle displacement components at the time $t$ = 0.1 s. In Figs. 5a and 5b, we highlight how the wavefield patterns change with a change in the orientation of the fault plane, particularly the strike of fault. If we have sufficient coverage of receivers around the source, this information could be helpful in focal mechanism determination, i.e., in finding the geometry of the fault and direction of slip on the fault plane. The compression and rarefaction patterns are opposite in Figs. 5a and 5b. This is because the azimuth angles of the fault-normal are different in these two cases. This clearly indicates why we should not neglect taking into account the strike of the fault plane in anisotropic media (or even isotropic media) while performing the source inversion.

\subsection{Inversion}

A total of 48 receivers are used in 3 boreholes, with 16 receivers per borehole (Fig. 8). The position vector for the true source location is (250, 200, 250) m. The initial guess is at (220, 210, 220) m. The three downhole array of sensors are at (x = 150, y = 350) m, (x = 350, y = 250) m, and (x = 350, y = 350) m. The spacing between consecutive sensors in an array is 30 m. 

In the local optimization algorithm, we have to begin with a close enough guess for the source parameters for it to converge. For example, using the traveltimes of microseismic waves, we can provide a good initial guess for the source position (see, for example, Grechka and Yaskevich, 2013). Next, we need to estimate the slip vector. The slip perturbation is related to the fault plane stiffness model \citep{kak}. The maximum amplitude of the recorded particle displacements is proportional to the slip on the fault plane. A rough estimate of the slip magnitude can be made from the above relations. The directivity of double-couple microseismic events (see Eisner and Stanek, 2018) is useful in approximating the slip angle and the dip angle of the fault. 

The initial guesses for the slip angle, the dip, the azimuth angle of the fault-normal, and the magnitude of the slip are $\pi/4$ rad, $\pi/5$ rad, $\pi/5$ rad and 0.5 m, respectively. Note that the guesses are not very close to the true source parameters. The waveform data (${\bf d}$) recorded at receivers corresponding to the total simulation time of 120 ms are shown in Fig. 6. 
\begin{figure}
\vspace*{-0.5in}
\centering
\begin{minipage}{1.0\textwidth}
\centering
    \includegraphics[width=3.5in]{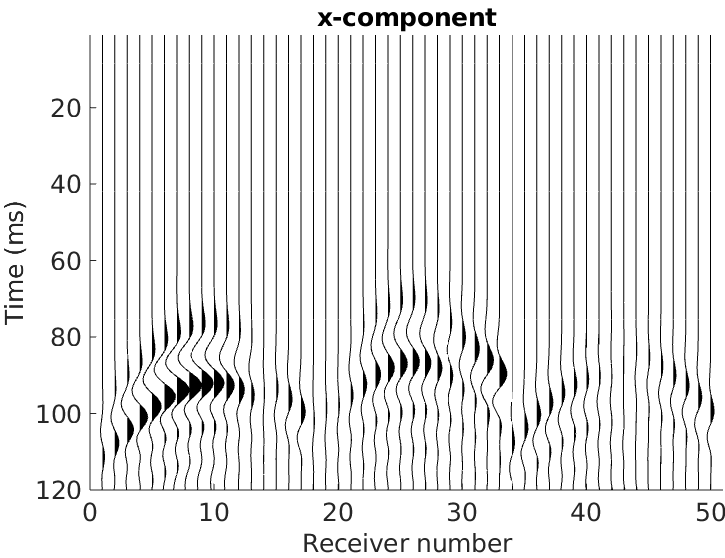}
\end{minipage}
\begin{minipage}{1.0\textwidth}
\centering
    \includegraphics[width=3.5in]{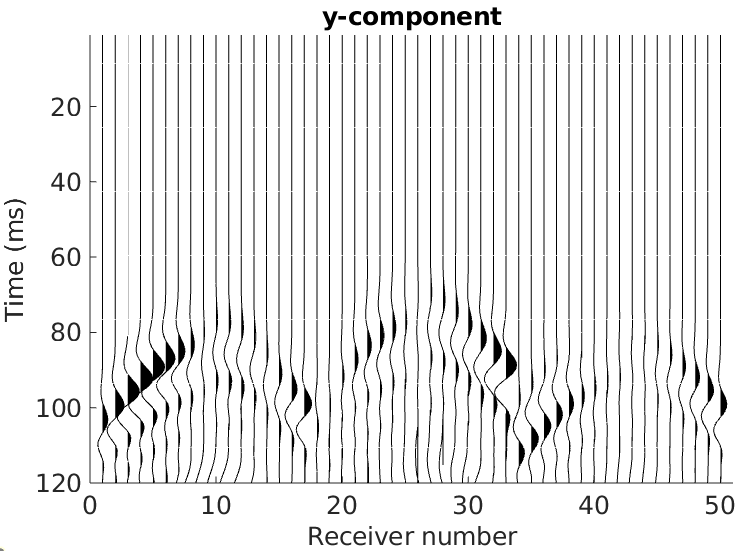}
\end{minipage}
\begin{minipage}{1.0\textwidth}
\centering
    \includegraphics[width=3.5in]{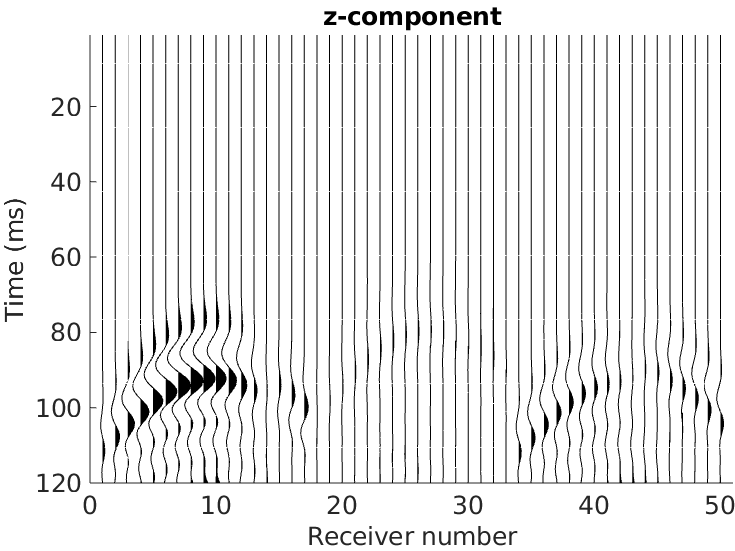}
\end{minipage}
\caption{The observed waveform data for the true double-couple source.}
\end{figure}

The calculated waveform data (${\bf u}$) corresponding to the same simulation time for the guess source are shown in Fig. 7.
\begin{figure}
\vspace*{-0.5in}
\centering
\begin{minipage}{1.0\textwidth}
\centering
    \includegraphics[width=3.5in]{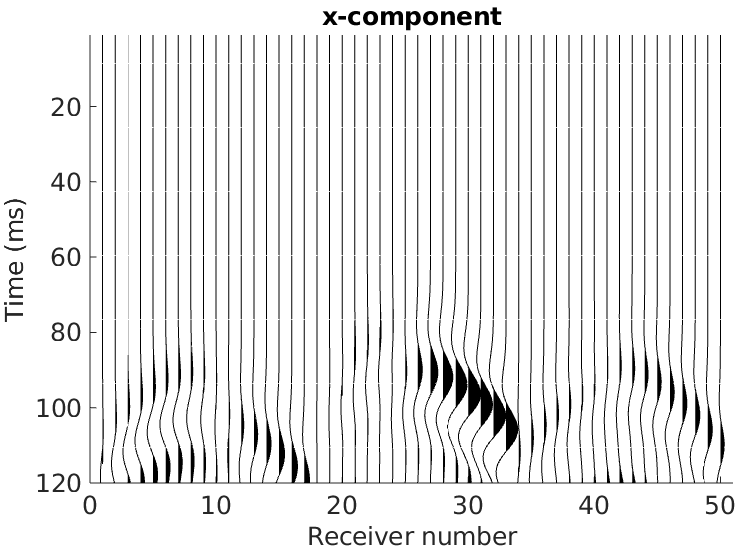}
\end{minipage}
\begin{minipage}{1.0\textwidth}
\centering
    \includegraphics[width=3.5in]{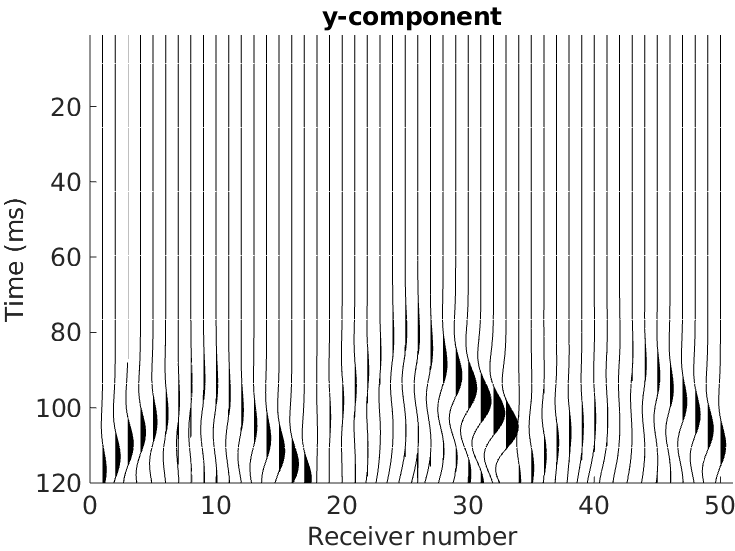}
\end{minipage}
\begin{minipage}{1.0\textwidth}
\centering
    \includegraphics[width=3.5in]{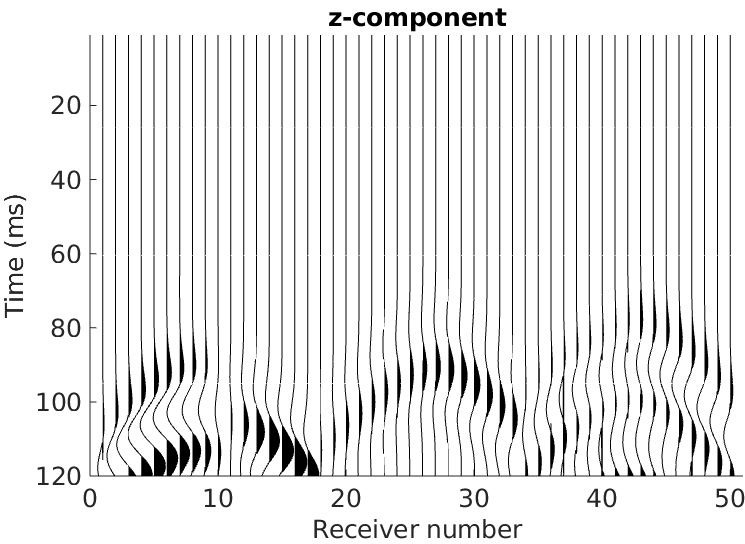}
\end{minipage}
\caption{The simulated waveform data for the guess double-couple source.}
\end{figure}

The iterative updates in the source position are demonstrated in Fig. 8 and the numerical values of the updated positions are given in Table 2. 
\begin{figure}
    \centering
    \includegraphics[width=0.8\textwidth]{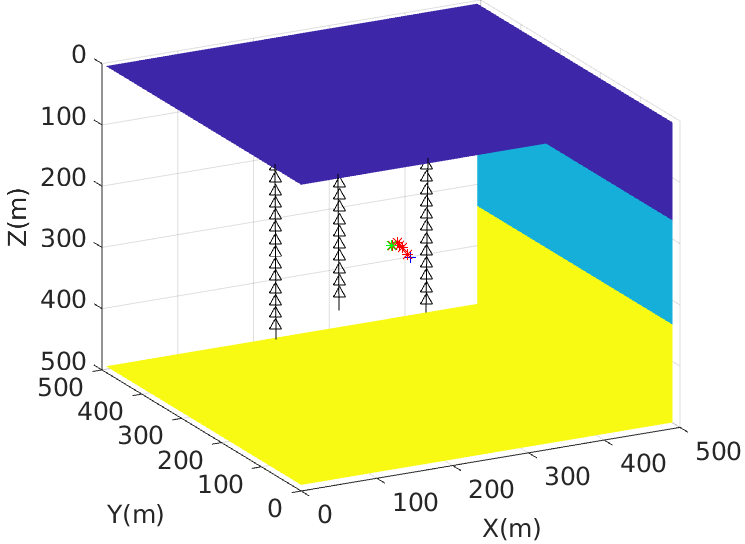}
    \caption{Iterative updates (\textcolor{red}{$*$}) in the source position in the double-couple source inversion. The true source location, the initial guess location and receivers are represented as \textcolor{blue}{+}, \textcolor{green}{$*$} and \textcolor{black}{$\Delta$} respectively. The updated positions gradually moved from the guess location to the true location.}
\end{figure}

\begin{table}
\begin{minipage}{175mm}
\caption{Updates in the inverted source location in the layered model with inversion iterations.}
\label{anymode}
\begin{tabular}{@{}llllllllllllll}
\hline
\hline
Iteration number & 0 & 2 & 4 & 6 & 8 & 10 & 12 & 14 & 16 & 18 & 20 & 22 & 24\\
\hline 
X (m) & 220 & 222 & 224 & 226 & 230 & 237 & 242 & 245 & 245 & 245 & 247 & 247 & 247\\[2pt] Y (m) & 210 & 207 & 206 & 203 & 204 & 202 & 200 & 201 & 202 & 202 & 202 & 202 & 202\\[2pt] Z (m) & 220 & 222 & 225 & 225 & 225 & 228 & 234 & 244 & 253 & 253 & 252 & 252 & 252\\[2pt]

\hline
\hline
\end{tabular}
\end{minipage}
\end{table}

It took 24 inversion iterations to minimize the objective function to the prescribed tolerance limit and also to bring the inverted source position vector close to the true source location (Fig. 9).
\begin{figure}
    \centering
    \includegraphics[width=0.6\textwidth]{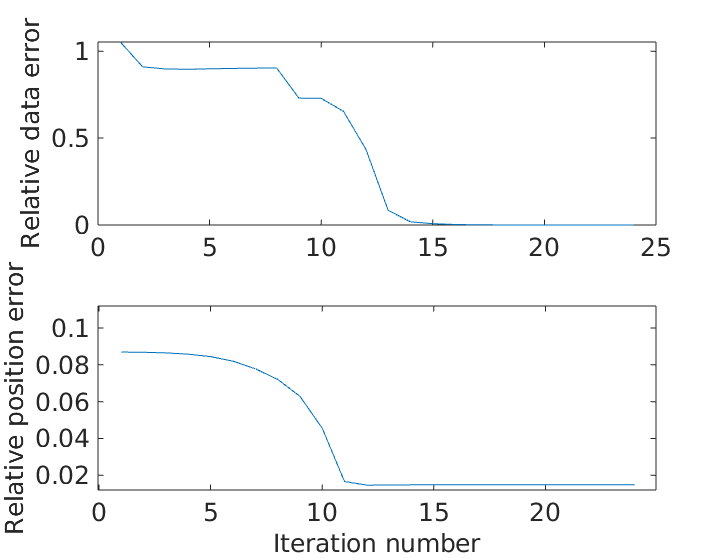}
    \caption{Plots of relative error (see Eqns. 13 and 23) between the observed and modelled data (above); and the true and updated source locations (below) with each inversion iteration.}
\end{figure}
All geometrical parameters are well recovered (Fig. 10). 
\begin{figure}
    \centering
    \includegraphics[width=0.8\textwidth]{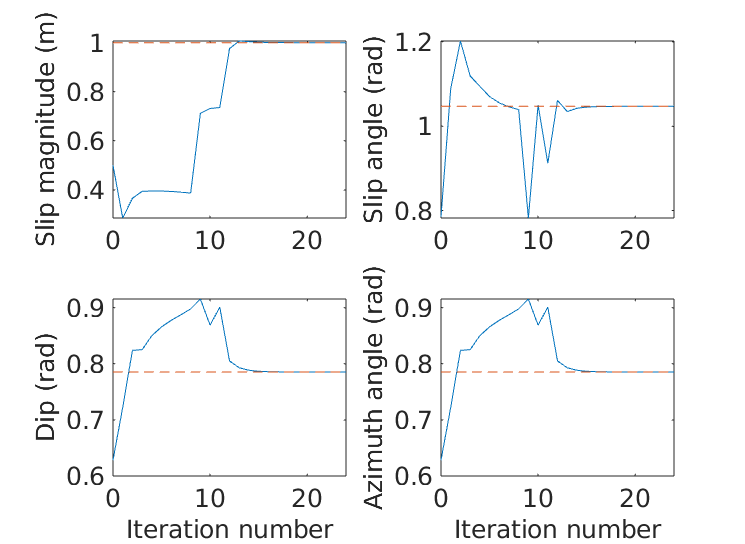}
    \caption{Comparison between the true (dashed lines) and modelled (solid lines) geometrical parameters with each inversion iteration.}
\end{figure}
The difference in the observed waveform data and the waveform data calculated using the inverted source parameters is negligible (Fig. 11).
\begin{figure}
\vspace*{-0.5in}
\centering
\begin{minipage}{1.0\textwidth}
\centering
    \includegraphics[width=3.5in]{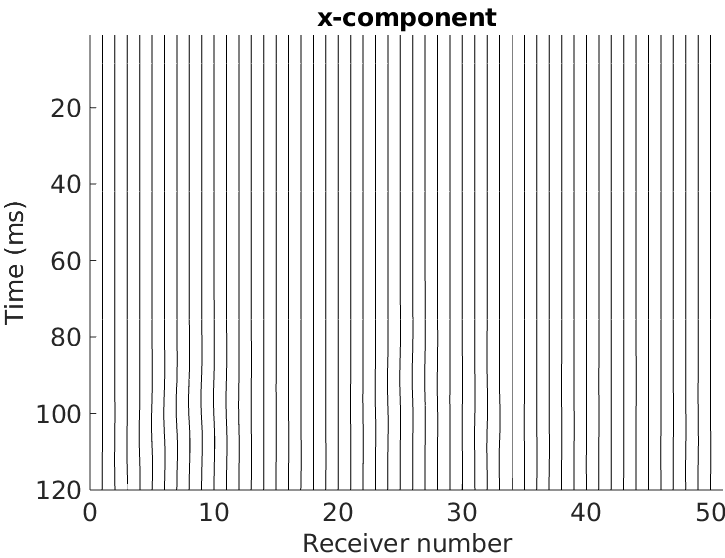}
\end{minipage}
\begin{minipage}{1.0\textwidth}
\centering
    \includegraphics[width=3.5in]{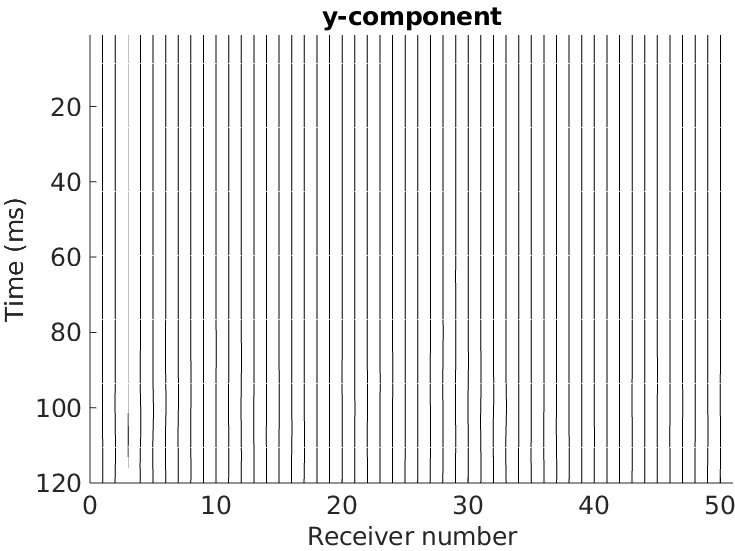}
\end{minipage}
\begin{minipage}{1.0\textwidth}
\centering
    \includegraphics[width=3.5in]{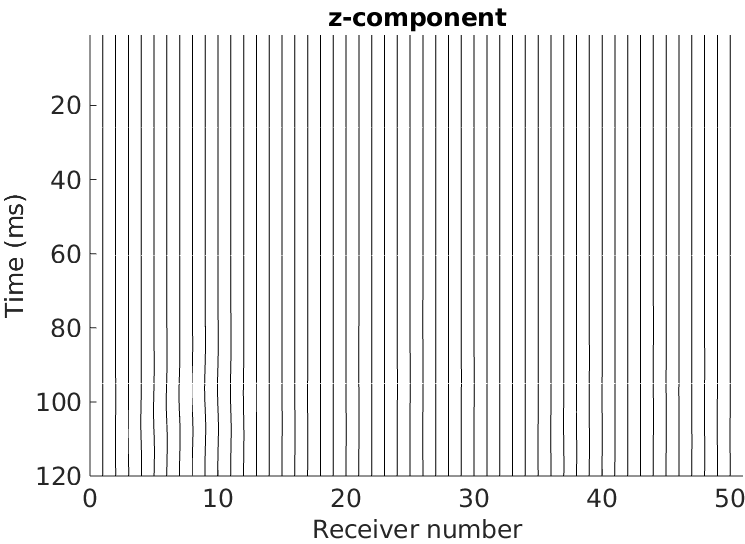}
\end{minipage}
\caption{The difference between the observed wiggle trace for the true double-couple source and the calculated wiggle trace for the inverted double-couple source.}
\end{figure}

\subsubsection{Limited coverage effects}

It is theoretically evident that if some receivers are not in the significant energy distribution region of the wavefield, then it will be difficult to constrain geometrical parameters from the waveform data. If the source is out of the coverage area, it is possible that the inversion scheme will fail. In one scenario, we reduced the number of boreholes from 3 to 2. This led to an increased number of inversion iterations for the initial guess and the true source parameters to converge. In another scenario, we reduced the number of boreholes from 3 to 1. The error between the observed and simulated data was never reduced to the tolerance limit in that case. We did not obtain accurate source parameters. 

\subsubsection{Inversion from noisy amplitude}

An additive white Gaussian noise (AWGN) is added to the observed data at each receiver. AWGN  mimics the effect of many random processes that occur in nature. The signal-to-noise ratio (SNR) is kept at 10. Normally the noisy amplitude data should distort the moment tensor, however, it also causes a small error in source localization. This is because the gradient for the source position depends on the elements of the moment tensor (Michel and Tsvankin, 2015). The main result of the study on the influence of noise on inversion coincides with the inference drawn from the similar analysis performed by Michel and Tsvankin (2015), i.e., off-diagonal elements ($M_{12}$, $M_{23}$) of the moment tensor are more distorted compared to the diagonal elements. In terms of geometrical parameters, it is the slip angle and the azimuth angle that are worse recovered. For the model considered, the minimum SNR that the algorithm can tolerate is 5. However, microseismic events recorded in a downhole experiment are usually less contaminated with noise. Furthermore, the downhole microseismic signal-to-noise ratio can be enhanced through the strip-matching shearlet transform \citep{LiJ}. The results of adding minimal noise to the observed data are similar to performing the source inversion in the model generated using seismic FWI instead of using the actual model. This is presented in the next subsection.

\subsubsection{Replacing the actual model with the FWI-generated model}

We want to test the quality of the inversion results when the true velocity model is not used. Therefore, we used a heterogeneous VFTI model that is generated through the cross-hole seismic FWI of the controlled-source elastic waveform data (Shekhar et al., 2025). The three fracture weaknesses, $\delta_N$, $\delta_V$ and $\delta_H$, can be determined through the cross-hole seismic FWI if the elastic parameters of the VTI background are known \citep{shekh}. The source inversion setup (Fig. 12) is the same as the previous setup shown in Fig. 8.
\begin{figure}
    \centering
    \includegraphics[width=0.8\textwidth]{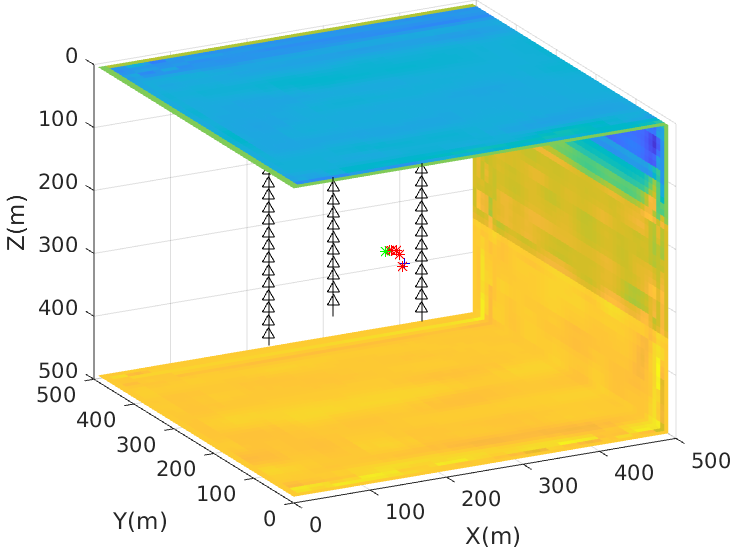}
    \caption{Iterative updates (\textcolor{red}{$*$}) in the source position when the actual layered model is replaced by the FWI-generated model. The true source location, the initial guess location and receivers are represented as \textcolor{blue}{+}, \textcolor{green}{$*$} and \textcolor{black}{$\Delta$} respectively. The inverted source position gradually moved from the guess position towards the true position.}
\end{figure}
\begin{figure}
    \centering
    \includegraphics[width=0.6\textwidth]{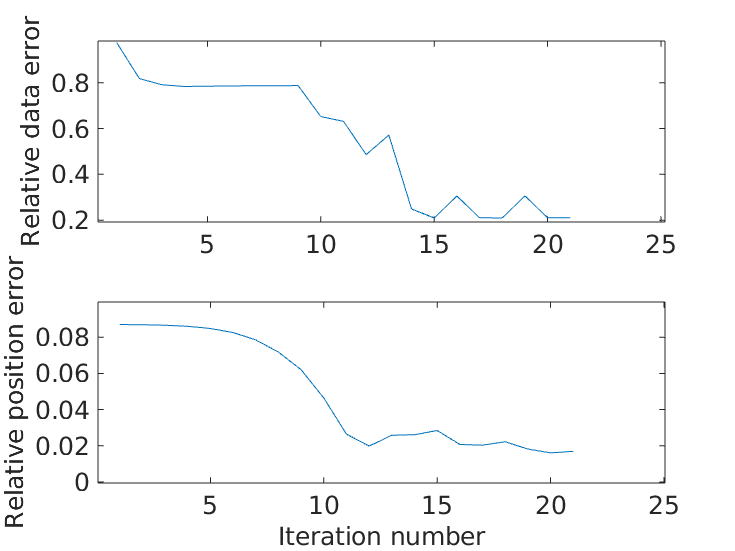}
    \caption{Plots of relative error between the observed and modelled data (above); and the true and updated source locations (below) with each inversion iteration when the actual layered model is replaced by the FWI-generated model.}
\end{figure}
\begin{figure}
    \centering
    \includegraphics[width=0.8\textwidth]{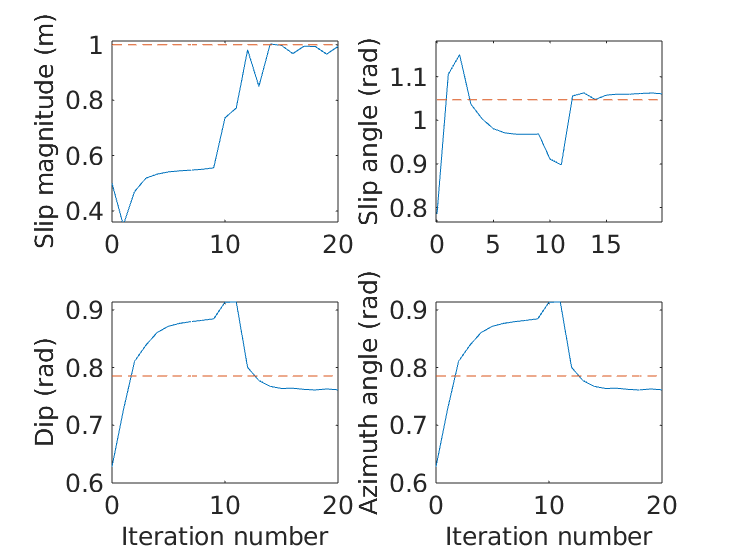}
    \caption{Comparison between the true (dashed lines) and modelled (solid lines) geometrical parameters with each inversion iteration when the actual layered model is replaced by the FWI-generated model.}
\end{figure}
\begin{figure*}
\vspace*{-0.5in}
\centering
\begin{minipage}{1.0\textwidth}
\centering
    \includegraphics[width=3.5in]{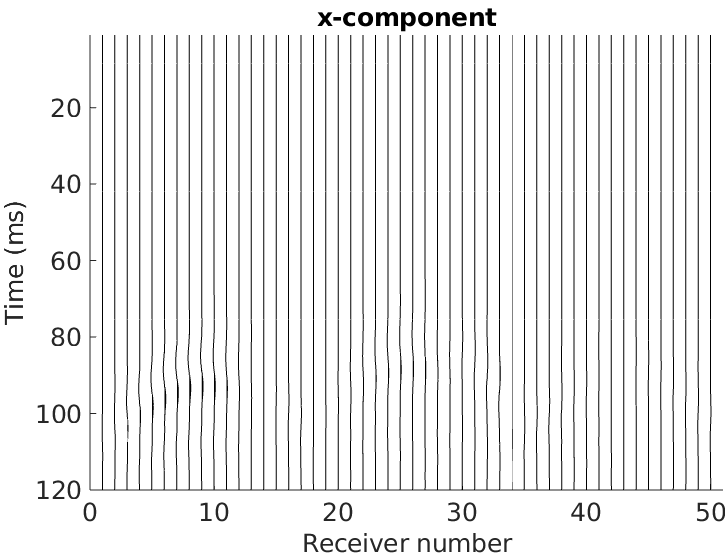}
\end{minipage}
\begin{minipage}{1.0\textwidth}
\centering
    \includegraphics[width=3.5in]{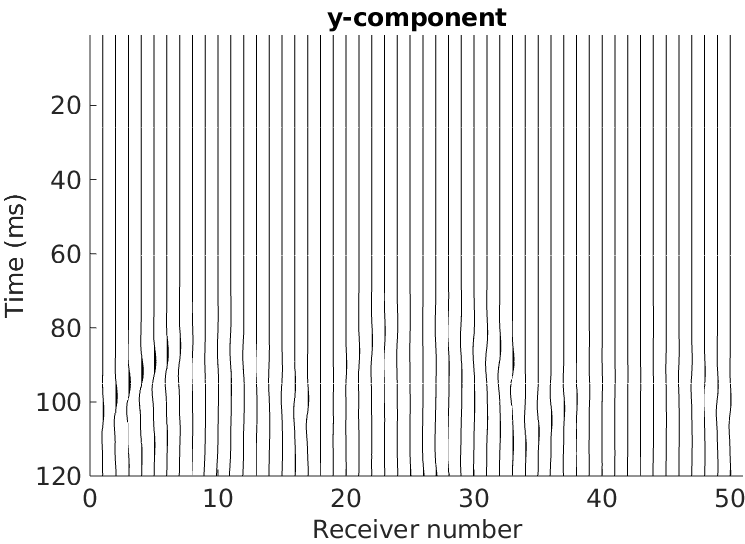}
\end{minipage}
\begin{minipage}{1.0\textwidth}
\centering
    \includegraphics[width=3.5in]{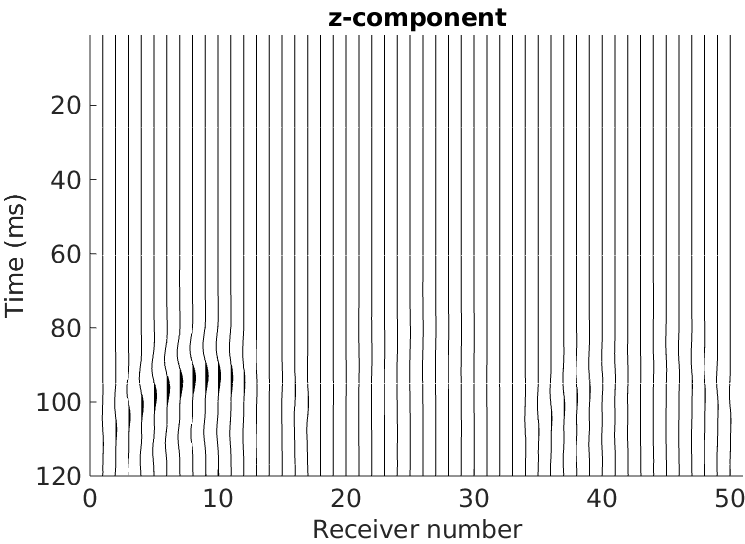}
\end{minipage}
\caption{The difference between the observed wiggle trace for the true double-couple source in the actual layered model and the calculated wiggle trace for the inverted double-couple source in the FWI-generated model.}
\end{figure*}

The guess source parameters are the same as described in Section 5.2.2. When using the FWI-generated model, the inverted parameters are close to the true parameters, but the relative data errors have not been minimized to a satisfactory limit (Fig. 13). The inversion scheme was unable to minimize the objective function after 21 iterations. The obtained position vector of the source is close enough to the true source location (see the value of the relative position error in Fig. 13). The magnitude and angle of the slip are well recovered (Fig. 14); however, there are minor errors in the inverted dip and azimuth angle (Fig. 14). The difference between the observed waveforms and the waveforms simulated using the inverted parameters for the source is shown in Fig. 15.
In general, the S-wave amplitudes have some mismatch for all the components of the particle displacement. However, this mismatch is highest in the z-(vertical) component. 
\begin{figure*}
    \centering
    \includegraphics[width=0.6\textwidth]{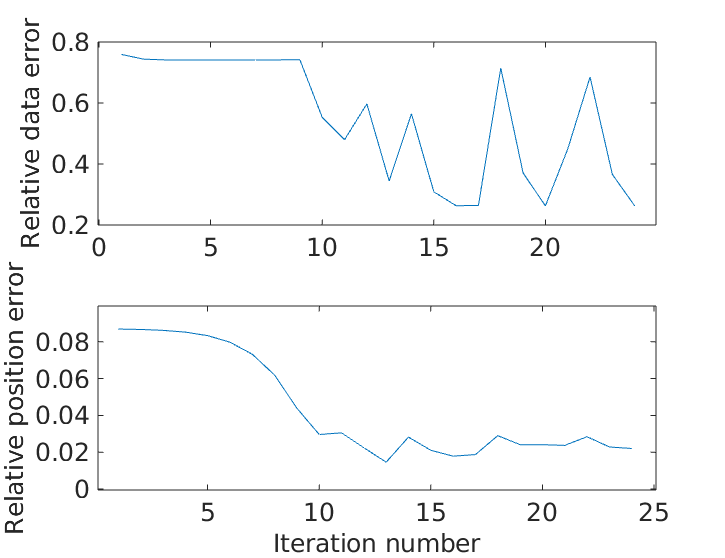}
    \caption{Plots of relative error between the observed and modelled data (above); and true and updated source location (below) with each inversion iteration when using the fracture-free model instead of the actual layered model.}
\end{figure*}
\begin{figure*}
    \centering
    \includegraphics[width=0.6\textwidth]{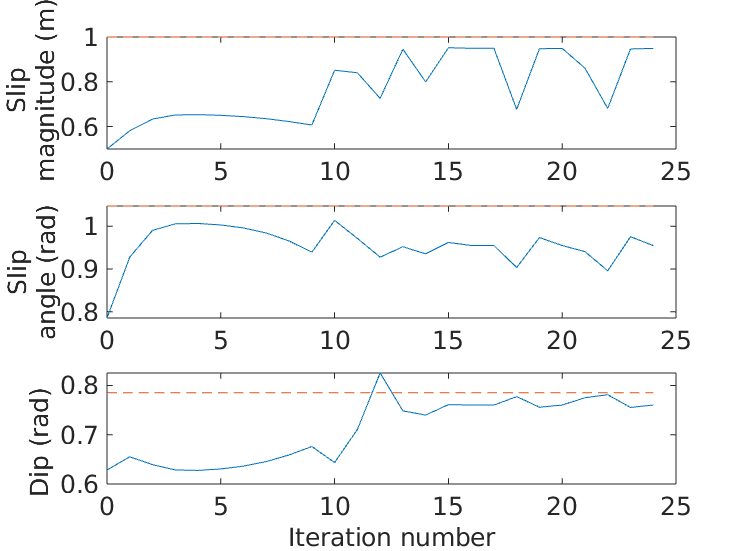}
    \caption{Comparison between the true (dashed lines) and modelled (solid lines) geometrical parameters with each inversion iteration when the actual layered (fractured) model is replaced by the fracture-free model.}
\end{figure*}
\begin{figure}
\vspace*{-0.5in}
\centering
\begin{minipage}{1.0\textwidth}
\centering
    \includegraphics[width=3.5in]{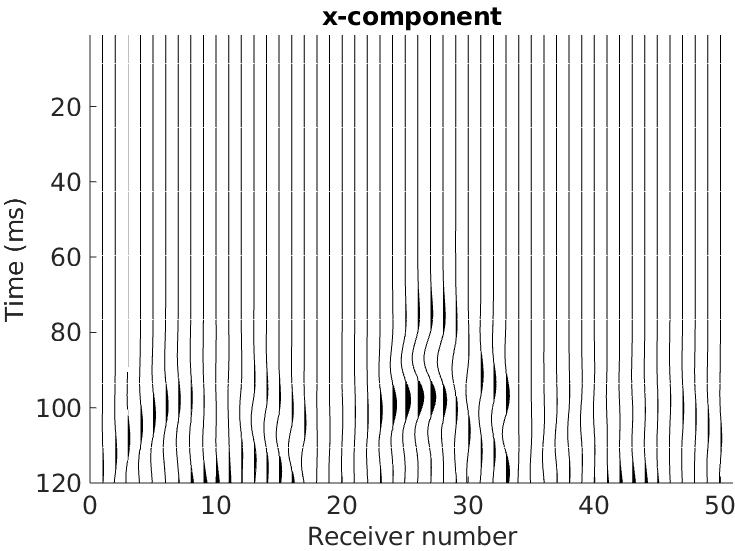}
\end{minipage}
\begin{minipage}{1.0\textwidth}
\centering
    \includegraphics[width=3.5in]{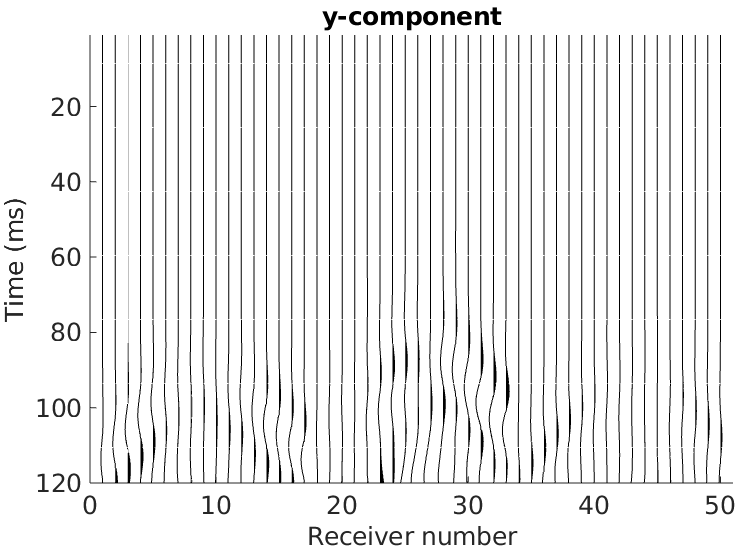}
\end{minipage}
\begin{minipage}{1.0\textwidth}
\centering
    \includegraphics[width=3.5in]{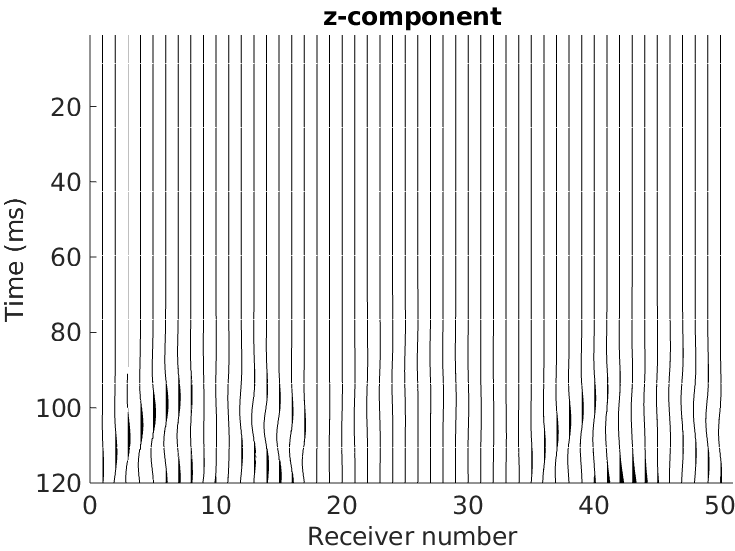}
\end{minipage}
\caption{The difference between the observed wiggle trace for the true double-couple source in the actual layered model and the calculated wiggle trace for the inverted double-couple source in the fracture-free (VTI layers) model.}
\end{figure}

\subsubsection{Replacing the actual model with the fracture-free model}

In this case, the velocity model is misinterpreted as being for a domain without vertical fractures and consists of three VTI layers. The model dimension and the thicknesses of the layers are the same as the dimension and thicknesses of the respective layers for the actual model. The elastic parameters of these layers are also the same as the elastic parameters $c^{(b)}_{IJ}$ of the unfractured VTI background for the actual model (see Table 1). The density values in the top, middle and bottom layers are 2250 kg/m$^3$, 2300 kg/m$^3$ and 2350 kg/m$^3$, respectively. All guess source parameters are the same as in Section 5.2.2 except the azimuth angle, which is fixed to zero rad. This means that the fault-normal lies on the xz-plane. The expression for the moment tensor of the double-couple source in VTI media is given in Appendix A. In this incorrect velocity model, the inversion is highly dependent on the frequency content of the source. If the dominant frequency of the source is 30 Hz, the inverted parameters are close to the true parameters. If the dominant frequency is on the order of 50 to 100 Hz, the numerical scheme fails to invert accurately. The relative data error is not reduced smoothly (Fig. 16). Even if the relative position error for the inverted source is low, we cannot trust the values for the position vector as long as the objective function is not minimized to the tolerance limit. The inverted dip is satisfactory, however, the inverted magnitude and angle of slip are not good enough (Fig. 17). It is important to note here that for a higher peak frequency source-characteristic, fracture-induced anisotropy must be taken into account. The difference between the observed waveforms and the waveforms simulated using the inverted parameters for the source is shown in Fig. 18. Both the P and S wave amplitudes have a significant mismatch for all the components of the particle displacement. However, this mismatch is highest in the x-component.

\section{Concluding remarks}

We have presented an elastic wave equation-based waveform inversion methodology to determine the location and moment tensor of a double-couple microseismic source in heterogeneous vertically fractured transversely isotropic (VFTI) media. Vertical fractures are commonly present in sedimentary rocks because of the triaxial stress-state within the Earth's crust. Neglecting such fractures leads to an incorrect estimate of the source location and the moment tensor. This is because fractures affect both the radiation pattern and the anisotropic focal mechanism of the double-couple source. To evaluate the moment tensor of the double-couple source, we need to know the geometry of the fault plane on which the slip has taken place. We do not restrict the orientation of the fault plane to any of the symmetry planes of the VFTI media, and it can have all possible dip and strike values. The moment tensor is therefore parameterized in terms of the geometrical parameters, which are the slip magnitude, slip angle, fault dip, and azimuth angle of the fault-normal. The Gauss-Newton method used in the local optimization employs the approximate Hessian parameter, which mitigates the cross-talk between different source parameters. In addition, incorporating the regularization parameter in the Gauss-Newton method helps the algorithm to converge. Inversion for four geometrical parameters instead of six moment-tensor components from waveform data in 3D anisotropic media reduces the computational cost involved. Geometrical parameters also provide information that is relevant from the geotechnical and geological point of view. We also noticed that during local optimization, we do not necessarily need close enough guesses for these parameters.

The densities, the background elastic parameters and the fracture weaknesses of the VFTI media are assumed to be known in the synthetic experiment involving the layered model. However, practically it is not possible to have exact values of these physical properties in the subsurface. Therefore, we employ a fracture model derived using the full waveform inversion of controlled-source seismic data. It is found that the relative data residual is not minimized after a certain iteration. The inverted source location and the moment tensor are still close to the true location and the true moment tensor, respectively. The quality of FWI-generated model is an important factor in getting good source parameters. Therefore, cross-hole seismic full waveform inversion is needed to have a better reconstruction of the target region. When we use a model in which fractures are completely neglected, neither the inverted source location nor the inverted moment tensor is correct. In terms of geometric elements, the dip of the fault plane is recovered better than the slip magnitude and the slip angle in that case. The azimuth angle of the fault-normal is the least well-recovered parameter whenever the actual model is not used. This suggests that the elastic waveform data are quite sensitive to the strike of the fault.

Acquisition geometry plays a crucial role in double-couple source inversion within heterogeneous VFTI media. Borehole receivers are preferred to surface receivers because of their higher signal-to-noise ratio. Moreover, for a fixed time interval, borehole receivers capture fewer multiple reflected and refracted waves compared to surface receivers. A greater number of direct wave recordings improves the accuracy of source parameter estimation. Seismic wave propagation is influenced by factors such as attenuation, velocity dispersion, and pore fluid flow. Future work could be directed to take these factors into account.

\section*{Acknowledgements}

The authors acknowledge the VISTA program, The Norwegian Academy of Science and Letters and Equinor, for their financial support of this project. U. Shekhar thanks Ivan Pšenčík for helpful suggestions.

\section*{Data availability}
The synthetic dataset can be obtained from the authors.

\newpage
\section*{Appendix}
\subsection*{Appendix A: Double-couple moment tensor in VTI media}

If we assume that no vertical fractures are present, the VFTI medium degenerates into the VTI medium. Due to the rotational symmetry of VTI media, we consider that the fault-normal $\hat{n}$ lie in the xz-plane. We assume that we know the azimuth direction of the fault-normal, and it coincides with x-axis. Then the elements of the moment tensor ${\bf M}^{(b)}$ of the double-couple source in VTI media is given as (Grechka, 2020)
 \begin{equation}
 \begin{aligned}
M^{(b)}_{11} &= l\frac{c^{(b)}_{11}-c^{(b)}_{13}}{2} \sin2\alpha \cos\varphi,\\
M^{(b)}_{12} &= lc^{(b)}_{66} \sin\alpha \sin\varphi,\\
M^{(b)}_{13} &= lc^{(b)}_{55} \cos2\alpha \cos\varphi,\\
M^{(b)}_{22} &= l\frac{c^{(b)}_{12}-c^{(b)}_{13}}{2} \sin2\alpha \cos\varphi,\\
M^{(b)}_{23} &= lc^{(b)}_{55} \cos\alpha \sin\varphi,\\
M^{(b)}_{33} &= l\frac{c^{(b)}_{13}-c^{(b)}_{33}}{2} \sin2\alpha \cos\varphi. 
\end{aligned}
\end{equation}

\newpage


\begin{thebibliography}{62}
\providecommand{\natexlab}[1]{#1}
\providecommand{\url}[1]{\texttt{#1}}
\expandafter\ifx\csname urlstyle\endcsname\relax
  \providecommand{\doi}[1]{doi: #1}\else
  \providecommand{\doi}{doi: \begingroup \urlstyle{rm}\Url}\fi

\bibitem[Aki and Richards(2002)]{aki}
K.~Aki and P.~G. Richards.
\newblock \emph{Quantitative Seismology}.
\newblock University Science Books, Mill Valley, California, 2002.

\bibitem[Alkhalifah(2016)]{alkh}
T.~A. Alkhalifah.
\newblock \emph{Full Waveform Inversion in an Anisotropic World. Where are the parameters hiding?}
\newblock EAGE Publications bv, 2016.

\bibitem[Artman et~al.(2010)Artman, Podladtchikov, and Witten]{art}
B.~Artman, I.~Podladtchikov, and B.~Witten.
\newblock Source location using time-reverse imaging.
\newblock \emph{Geophysical Prospecting}, 58\penalty0 (5):\penalty0 861--873, 2010.

\bibitem[Aster et~al.(2019)Aster, Borchers, and Thurber]{ast}
R.~C. Aster, B.~Borchers, and C.~H. Thurber.
\newblock \emph{Parameter Estimation and Inverse Problems}.
\newblock Elsevier, 2019.

\bibitem[Auld(1990)]{auld}
B.~Auld.
\newblock \emph{Acoustic Fields and Waves in Solids}.
\newblock Krieger Publishing Company, 1990.

\bibitem[Boitz et~al.(2018)Boitz, Reshetnikov, and Shapiro]{boi}
N.~Boitz, A.~Reshetnikov, and S.~A. Shapiro.
\newblock Visualizing effects of anisotropy on seismic moments and their potency-tensor isotropic equivalent.
\newblock \emph{Geophysics}, 83\penalty0 (3):\penalty0 C85--C97, 2018.

\bibitem[Brossier et~al.(2009)Brossier, Operto, and Virieux]{bro}
R.~Brossier, S.~Operto, and J.~Virieux.
\newblock Seismic imaging of complex onshore structures by 2d elastic frequency-domain full-waveform inversion.
\newblock \emph{Geophysics}, 74\penalty0 (6):\penalty0 WCC105--WCC118, 2009.

\bibitem[Brune(1970)]{bru}
J.~Brune.
\newblock Tectonic stress and the spectra of seismic shear waves from earthquakes.
\newblock \emph{Journal of Geophysical Research}, 75:\penalty0 4997--5009, 1970.

\bibitem[Chambers et~al.(2010)Chambers, Kendall, and Barkved]{cham}
K.~Chambers, J.~Kendall, and O.~Barkved.
\newblock Investigation of induced microseismicity at valhall using the life of field seismic array.
\newblock \emph{The Leading Edge}, 29\penalty0 (3):\penalty0 290--295, 2010.

\bibitem[Choi and Alkhalifah(2011)]{choi}
Y.~Choi and T.~Alkhalifah.
\newblock Source-independent time-domain waveform inversion using convolved wavefields: application to the encoded multisource waveform inversion.
\newblock \emph{Geophysics}, 76\penalty0 (5):\penalty0 R125--R134, 2011.

\bibitem[Duncan and Eisner(2010)]{dun}
P.~M. Duncan and L.~Eisner.
\newblock Reservoir characterization using surface microseismic monitoring.
\newblock \emph{Geophysics}, 75\penalty0 (5):\penalty0 A139--A146, 2010.

\bibitem[Eaton(2008)]{eat}
D.~W. Eaton.
\newblock Microseismic focal mechanisms: A tutorial.
\newblock \emph{CREWES Research Report}, 20, 2008.

\bibitem[Eaton and Forouhideh(2011)]{eatfor}
D.~W. Eaton and F.~Forouhideh.
\newblock Solid angles and the impact of receiver-array geometry on microseismic moment-tensor inversion.
\newblock \emph{Geophysics}, 76\penalty0 (6):\penalty0 WC77--WC85, 2011.

\bibitem[Eisner and Stanek(2018)]{es}
L.~Eisner and F.~Stanek.
\newblock Microseismic data interpretation — what do we need to measure first?
\newblock \emph{First Break}, 36\penalty0 (2):\penalty0 55--58, 2018.

\bibitem[Gajewski and Tessmer(2005)]{gaj}
D.~Gajewski and E.~Tessmer.
\newblock Reverse modelling for seismic event characterization.
\newblock \emph{Geophysical Journal International}, 163\penalty0 (1):\penalty0 276--284, 2005.

\bibitem[Geiger(1912)]{gei}
L.~Geiger.
\newblock Probability method for the determination of earthquake epicenters from the arrival time only.
\newblock \emph{Bulletin of St. Louis University}, 8:\penalty0 60--71, 1912.

\bibitem[Gibson and Ben-Menahem(1991)]{gib}
R.~L. Gibson and A.~Ben-Menahem.
\newblock Elastic wave scattering by anisotropic obstacles: Application to fractured volumes.
\newblock \emph{Journal of Geophysical Research}, 96\penalty0 (B12), 1991.

\bibitem[Grechka(2020)]{gre}
V.~Grechka.
\newblock Moment tensors of double-couple microseismic sources in anisotropic formations.
\newblock \emph{Geophysics}, 85\penalty0 (1):\penalty0 KS1--KS11, 2020.

\bibitem[Grechka and Yaskevich(2013)]{yas}
V.~Grechka and S.~Yaskevich.
\newblock Inversion of microseismic data for triclinic velocity models.
\newblock \emph{Geophysical Prospecting}, 61\penalty0 (6):\penalty0 1159--1170, 2013.

\bibitem[Grechka and Yaskevich(2014)]{greyas}
V.~Grechka and S.~Yaskevich.
\newblock Azimuthal anisotropy in microseismic monitoring: A bakken case study.
\newblock \emph{Geophysics}, 79\penalty0 (1):\penalty0 KS1--KS12, 2014.

\bibitem[Huang et~al.(2017)Huang, Dong, Liu, and Yang]{hua}
C.~Huang, L.~Dong, Y.~Liu, and J.~Yang.
\newblock Acoustic wave-equation based full-waveform microseismic source location using improved scattering-integral approach.
\newblock \emph{Geophysical Journal International}, 209\penalty0 (3):\penalty0 1476--1488, 2017.

\bibitem[Jakobsen et~al.(2023)Jakobsen, Xiang, and van Dongen]{jako}
M.~Jakobsen, K.~Xiang, and K.~van Dongen.
\newblock Seismic and medical ultrasound imaging of velocity and density variations by nonlinear vectorial inverse scattering.
\newblock \emph{The Journal of the Acoustical Society of America}, 153\penalty0 (5):\penalty0 3151--3164, 2023.

\bibitem[Kakurina et~al.(2019)Kakurina, Guglielmi, Nussbaum, and Valley]{kak}
M.~Kakurina, Y.~Guglielmi, C.~Nussbaum, and B.~Valley.
\newblock Slip perturbation during fault reactivation by a fluid injection.
\newblock \emph{Tectonophysics}, 757:\penalty0 140--152, 2019.

\bibitem[Kamei and Lumley(2017)]{kam}
R.~Kamei and D.~Lumley.
\newblock Full waveform inversion of repeating seismic events to estimate time-lapse velocity changes.
\newblock \emph{Geophysical Journal International}, 209\penalty0 (2):\penalty0 1239--1264, 2017.

\bibitem[Kim et~al.(2011)Kim, Liu, and Tromp]{kim}
Y.~Kim, Q.~Liu, and J.~Tromp.
\newblock Adjoint centroid-moment tensor inversions.
\newblock \emph{Geophysical Journal International}, 186\penalty0 (1):\penalty0 264--278, 2011.

\bibitem[Lei et~al.(2021)Lei, Tan, Zhang, Malkoti, Abakumov, and Xie]{lei}
L.~Lei, J.~Tan, D.~Zhang, A.~Malkoti, I.~Abakumov, and Y.~Xie.
\newblock Fdwave3d: a matlab solver for the 3d anisotropic wave equation using the finite-difference method.
\newblock \emph{Computational Geosciences}, 25\penalty0 (5):\penalty0 1565--1578, 2021.

\bibitem[Li et~al.(2018)Li, Ji, Li, Qian, and Lu]{LiJ}
J.~Li, S.~Ji, Y.~Li, Z.~Qian, and W.~Lu.
\newblock Downhole microseismic signal-to-noise ratio enhancement via strip matching shearlet transform.
\newblock \emph{Journal of Geophysics and Engineering}, 15\penalty0 (2):\penalty0 330--337, 2018.

\bibitem[Li et~al.(2020)Li, Tan, Schwarz, Staněk, Poiata, Shi, Diekmann, Eisner, and Gajewski]{li}
L.~Li, J.~Tan, B.~Schwarz, F.~Staněk, N.~Poiata, P.~Shi, L.~Diekmann, L.~Eisner, and D.~Gajewski.
\newblock Recent advances and challenges of waveform-based seismic location methods at multiple scales.
\newblock \emph{Reviews of Geophysics}, 58\penalty0 (1), 2020.

\bibitem[Lions and Magenes(1972)]{lio}
J.~Lions and E.~Magenes.
\newblock \emph{Nonhomogeneous boundary value problems and applications}.
\newblock Grundlehren der mathematischen Wissenschaften, Springer, 1972.

\bibitem[Liu and Tromp(2006)]{liu}
Q.~Liu and J.~Tromp.
\newblock Finite-frequency kernels based on adjoint methods.
\newblock \emph{Bulletin of the Seismological Society of America}, 96\penalty0 (6):\penalty0 2383--2397, 2006.

\bibitem[Madariaga(2015)]{mad}
R.~Madariaga.
\newblock \emph{Seismic source theory}.
\newblock Elsevier B.V., 2015.

\bibitem[McMechan(1982)]{mcm}
G.~A. McMechan.
\newblock Determination of source parameters by wavefield extrapolation.
\newblock \emph{Geophysical Journal of the Royal Astronomical Society}, 71\penalty0 (3):\penalty0 613--628, 1982.

\bibitem[Menke(2012)]{men}
W.~Menke.
\newblock \emph{Geophysical data analysis: discrete inverse theory}.
\newblock Academic press, 2012.

\bibitem[Mewis and Wagner(2009)]{mew}
J.~Mewis and N.~J. Wagner.
\newblock Thixotropy.
\newblock \emph{Advances in Colloid and Interface Science}, 147-148:\penalty0 214--227, 2009.

\bibitem[Michel and Tsvankin(2015)]{tsv}
O.~J. Michel and I.~Tsvankin.
\newblock Estimation of microseismic source parameters by 2d anisotropic waveform inversion.
\newblock \emph{Journal of Seismic Exploration}, 24\penalty0 (4):\penalty0 379--400, 2015.

\bibitem[Michel and Tsvankin(2017)]{mic}
O.~J. Michel and I.~Tsvankin.
\newblock Waveform inversion for microseismic velocity analysis and event location in vti media.
\newblock \emph{Geophysics}, 82\penalty0 (4):\penalty0 WA95--WA103, 2017.

\bibitem[Michel and Tsvankin(2019)]{mich}
O.~J. Michel and I.~Tsvankin.
\newblock 3d waveform inversion of downhole microseismic data for transversely isotropic media.
\newblock \emph{Geophysical Prospecting}, 67\penalty0 (9):\penalty0 2332--2342, 2019.

\bibitem[Mora(1987)]{mor}
P.~Mora.
\newblock Nonlinear two-dimensional elastic inversion of multioffset seismic data.
\newblock \emph{Geophysics}, 52\penalty0 (9):\penalty0 1211--1228, 1987.

\bibitem[Morency and Mellors(2012)]{mm}
C.~Morency and R.~Mellors.
\newblock Full moment tensor and source location inversion based on full waveform adjoint inversion: application at the geysers geothermal field.
\newblock \emph{82$^{nd}$ Annual International Meeting, SEG, Expanded Abstracts}, pages 1--5, 2012.

\bibitem[Nocedal and Wright(2000)]{nw}
J.~Nocedal and S.~J. Wright.
\newblock \emph{Numerical Optimization}.
\newblock Springer Science$+$Business Media, LLC., 2000.

\bibitem[Oh and Alkhalifah(2019)]{oha}
J.~W. Oh and T.~Alkhalifah.
\newblock Study on the full-waveform inversion strategy for 3d elastic orthorhombic anisotropic media: application to ocean bottom cable data.
\newblock \emph{Geophysical prospecting}, 67\penalty0 (5):\penalty0 1219--1242, 2019.

\bibitem[Plessix(2006)]{ple}
R.~E. Plessix.
\newblock A review of the adjoint-state method for computing the gradient of a functional with geophysical applications.
\newblock \emph{Geophysical Journal International}, 167\penalty0 (2):\penalty0 495--503, 2006.

\bibitem[Rössler et~al.(2004)Rössler, Rümpker, and Krüger]{ros}
D.~Rössler, G.~Rümpker, and F.~Krüger.
\newblock Ambiguous moment tensors and radiation patterns in anisotropic media with applications to the modeling of earthquake mechanisms in w-bohemia.
\newblock \emph{Studia Geophysica et Geodaetica}, 48\penalty0 (1):\penalty0 233--250, 2004.

\bibitem[Schoenberg(1983)]{scho}
M.~Schoenberg.
\newblock Reflection of elastic waves from periodically stratified media with interfacial slip.
\newblock \emph{Geophysical Prospecting}, 31\penalty0 (2):\penalty0 265--292, 1983.

\bibitem[Schoenberg and Helbig(1997)]{sch}
M.~Schoenberg and K.~Helbig.
\newblock Orthorhombic media: Modeling elastic wave behavior in a vertically fractured earth.
\newblock \emph{Geophysics}, 62\penalty0 (6):\penalty0 1954--1974, 1997.

\bibitem[Shearer(1999)]{shea}
P.~M. Shearer.
\newblock \emph{Introduction to Seismology}.
\newblock Cambridge University Press, 1999.

\bibitem[Shekhar(2021)]{shek}
U.~Shekhar.
\newblock \emph{Effective seismic model from fractured rock}.
\newblock ntnuopen.ntnu.no/ntnu-xmlui/handle/11250/2834586, 2021.

\bibitem[Shekhar et~al.(2025)Shekhar, Jakobsen, Pšenčík, and Xiang]{shekh}
U.~Shekhar, M.~Jakobsen, I.~Pšenčík, and K.~Xiang.
\newblock Seismic full waveform inversion for fracture parameters in anisotropic media.
\newblock \emph{Geophysical Prospecting}, 73\penalty0 (5):\penalty0 1606--1634, 2025.

\bibitem[Shi et~al.(2018)Shi, Angus, Nowacki, Yuan, and Wang]{shi}
P.~Shi, D.~Angus, A.~Nowacki, S.~Yuan, and Y.~Wang.
\newblock Microseismic full waveform modeling in anisotropic media with moment tensor implementation.
\newblock \emph{Surveys in Geophysics}, 39\penalty0 (5):\penalty0 567--611, 2018.

\bibitem[Stein and Wysession(2003)]{sw}
S.~Stein and M.~Wysession.
\newblock \emph{An introduction to seismology, earthquakes, and earth structure}.
\newblock Blackwell publishing ltd, 2003.

\bibitem[Thomsen(1986)]{thom}
L.~Thomsen.
\newblock Weak elastic anisotropy.
\newblock \emph{Geophysics}, 51\penalty0 (10):\penalty0 1954--1966, 1986.

\bibitem[Tong et~al.(2016)Tong, Yang, Liu, Yang, and Harris]{tong}
P.~Tong, D.~Yang, Q.~Liu, X.~Yang, and J.~Harris.
\newblock Acoustic wave equation-based earthquake location.
\newblock \emph{Geophysical Journal International}, 205\penalty0 (1):\penalty0 464--478, 2016.

\bibitem[Tromp et~al.(2005)Tromp, Tape, and Liu]{trom}
J.~Tromp, C.~Tape, and Q.~Liu.
\newblock Seismic tomography, adjoint methods, time reversal and banana-doughnut kernels.
\newblock \emph{Geophysical Journal International}, 160\penalty0 (1):\penalty0 195--216, 2005.

\bibitem[Vavryčuk(2005)]{vav2005}
V.~Vavryčuk.
\newblock Focal mechanisms in anisotropic media.
\newblock \emph{Geophysical Journal International}, 161\penalty0 (2):\penalty0 334--346, 2005.

\bibitem[Vavryčuk(2007)]{vav2007}
V.~Vavryčuk.
\newblock On the retrieval of moment tensors from borehole data.
\newblock \emph{Geophysical Prospecting}, 55\penalty0 (3):\penalty0 381--391, 2007.

\bibitem[Vavryčuk(2015)]{vav2015}
V.~Vavryčuk.
\newblock Moment tensor decompositions revisited.
\newblock \emph{Journal of Seismology}, 19:\penalty0 231--252, 2015.

\bibitem[Virieux(1986)]{vir}
J.~Virieux.
\newblock P-sv wave propagation in heterogeneous media: Velocity-stress finite-difference method.
\newblock \emph{Geophysics}, 51\penalty0 (4):\penalty0 889--901, 1986.

\bibitem[Virieux and Operto(2009)]{vo}
J.~Virieux and S.~Operto.
\newblock An overview of full-waveform inversion in exploration geophysics.
\newblock \emph{Geophysics}, 74\penalty0 (6):\penalty0 WCC1--WCC26, 2009.

\bibitem[Wang and Alkhalifah(2018)]{wan}
H.~Wang and T.~Alkhalifah.
\newblock Microseismic imaging using a source function independent full waveform inversion method.
\newblock \emph{Geophysical Journal International}, 214\penalty0 (1):\penalty0 46--57, 2018.

\bibitem[Wang et~al.(2020)Wang, Guo, Alkhalifah, and Wu]{wang}
H.~Wang, Q.~Guo, T.~Alkhalifah, and Z.~Wu.
\newblock Regularized elastic passive equivalent source inversion with full-waveform inversion: Application to a field monitoring microseismic data set.
\newblock \emph{Geophysics}, 85\penalty0 (6):\penalty0 KS207--KS219, 2020.

\bibitem[Xiang et~al.(2024)Xiang, Jakobsen, Shekhar, Eikrem, and Nævdal]{xian}
K.~Xiang, M.~Jakobsen, U.~Shekhar, K.~S. Eikrem, and G.~Nævdal.
\newblock Efficient scattering approach to seismic full-waveform inversion in anisotropic elastic media with variable density.
\newblock \emph{arXiv:2402.02116}, 2024.

\bibitem[Šílený et~al.(1992)Šílený, Panza, and Campus]{sil}
J.~Šílený, G.~F. Panza, and P.~Campus.
\newblock Waveform inversion for point source moment tensor retrieval with variable hypocentral depth and structural model.
\newblock \emph{Geophysical Journal International}, 109\penalty0 (2):\penalty0 259--274, 1992.

\end{thebibliography}
\end{document}